\documentclass[11pt,epsfig,letterpaper]{article}
\usepackage{varioref,exscale,latexsym,amsmath,amssymb}
\usepackage{graphicx}
\def\beq{\begin{equation}}

\def\eeq{\end{equation}}

\def\beqa{\begin{eqnarray}}

\def\eeqa{\end{eqnarray}}

\begin{document}


\begin{titlepage}

\vspace*{-15mm}
\begin{flushright}
MPP-2009-57\\
\end{flushright}
\vspace*{3mm}

\begin{center}
{
\bf\LARGE
Non-isotropy in the CMB power spectrum in single field inflation
}
\\[8mm]
John F. Donoghue$^{a}$
\footnote{E-mail: \texttt{donoghue@physics.umass.edu}},
Koushik Dutta$^{a,b}$
\footnote{E-mail: \texttt{koushik@mppmu.mpg.de}},
Andreas Ross$^{a,c}$
\footnote{E-mail: \texttt{andreas.ross@yale.edu}},\\

\end{center}
\vspace*{0.50cm}
\centerline{$^a$  Department of Physics}
\centerline{ University of Massachusetts}
\centerline{Amherst, MA  01003, USA}

\vspace*{0.1cm}
\centerline{$^{b}$
Max-Planck-Institut f\"ur Physik (Werner-Heisenberg-Institut)}
\centerline{
F\"ohringer Ring 6, D-80805 M\"unchen, Germany}
\vspace*{0.1cm}

\vspace*{0.1cm}
\centerline{$^{c}$ Department of Physics}
\centerline {Yale University}
\centerline{ New Haven, CT 06520, USA}
\vspace*{0.1cm}

\begin{abstract}

\noindent
Contaldi $et~ al.$ \cite{Contaldi:2003zv} have suggested that an
initial period of kinetic energy domination in single field
inflation may explain the lack of CMB power at large angular scales.
We note that in this situation it is natural that there also be a
spatial gradient in the initial value of the inflaton field, and
that this can provide a spatial asymmetry in the observed CMB power
spectrum, manifest at low values of $\ell$. We investigate the
nature of this asymmetry and comment on its relation to possible
anomalies at low $\ell$.
\end{abstract}

\end{titlepage}

\newpage

\setcounter{footnote}{0}

\section{Introduction}
The simplest models of inflation involve only a single scalar
field, the inflaton \cite{inflation}. With a suitably chosen
potential, such a model can provide a simple explanation of the
temperature fluctuations in the CMB at all angular scales
\cite{Spergel:2006hy}. There are however a number of possible
anomalies in the CMB power spectrum that have attracted the
attention of many researchers \cite{Eriksen},
\cite{Hansen:2004vq}, \cite{north_south_non_gaussianity},
\cite{general_anomaly}, \cite{coldspot},
\cite{Eriksen:2007pc}, \cite{collection},
\cite{Hansen:2006rj}, \cite{Maino:2006pq}, \cite{Raeth:2007ti}, \cite{wmap5_evidence}.
(However see also \cite{Hajian}.) None of these anomalies is by
itself statistically compelling; however, taken together they
provide a hint that these features may be significant. Much
discussion of anomalies involves the power spectrum at low $\ell$,
$i.e.$ at large scales, where several anomalies indicate a
possible spatial asymmetry in the power spectrum, most often
roughly a north-south galactic coordinate asymmetry
\cite{Eriksen}, \cite{Hansen:2004vq}, \cite{north_south_non_gaussianity},
\cite{Eriksen:2007pc}, \cite{Hansen:2006rj}, \cite{Maino:2006pq}, \cite{Raeth:2007ti}, \cite{wmap5_evidence}.

The possibility that there may be an asymmetry in the observed CMB
power spectrum was first raised by Eriksen $et ~ al$
\cite{Eriksen} and Hansen $et ~ al$ \cite{Hansen:2004vq} using the
first year WMAP data. Their data analysis suggested a difference
in power of roughly 20\% for low $\ell$ maximized in the direction
of galactic coordinates $(80^\circ,57^\circ)$. Interestingly no
effect was seen above $\ell \sim 40$. For example, the analysis of
the power spectrum in the vicinity of the first acoustic peak
\cite{Hansen:2004vq, evan} showed no evidence of a spatial
asymmetry. At low values of $\ell$, the cosmic variance provides
an intrinsic scatter in the power spectrum data, so that even
though the signal is rather large, the statistical significance of
their result was below 3 standard deviations.

In their three year data release \cite{Spergel:2006hy}, the WMAP
team addressed the isotropy of the power spectrum, finding a small
asymmetry in a direction consistent with Eriksen $et ~ al$
\cite{Eriksen}. The method introduced by the WMAP team to
investigate asymmetries in the CMB spectrum is to multiply an
isotropic Gaussian CMB field by a large scale modulation function.
They test both a dipole and a quadrupole modulation and find that
the significance of the signal is not statistically compelling.
Their analysis uses a pixel size of $7^\circ$ which makes their
analysis sensitive up to $\ell \sim 20$. The original Eriksen team
has also revisited the WMAP 3 year data \cite{Eriksen:2007pc} using
a statistical framework similar to the WMAP team's with a modulation
function. They choose a higher resolution with a pixel size of
$3.6^\circ$ including multipoles up to $\ell \sim 40$ in their
analysis and confirm the asymmetry with a higher statistical
significance than the WMAP team and in consistency with their
previous analysis of the first year WMAP data.

Hansen $et~ al$ \cite{Hansen:2006rj} and Maino $et~ al$ \cite{Maino:2006pq}
explored two different approaches to extract the CMB spectrum where
WMAP data itself is used for foreground removal, and both find an
asymmetry of the power spectrum at largest scales consistent with
previous analysis and with each other. The fact that the asymmetry
does not vary when different foreground subtraction procedures
are applied constitutes a strong argument against a galactic origin
for the asymmetry. Moreover, the asymmetry was also found in
COBE data \cite{Eriksen} which indicates that systematics may not
be the correct explanation for a large scale asymmetry in the CMB
power spectrum. R\"ath $et ~ al$ \cite{Raeth:2007ti} have also found
the asymmetry in the WMAP 3 year data using statistical techniques
different from the ones used in previous analyses.
Recent analyses of WMAP 5 year data show a similar anisotropy of power between the two hemispheres, but with the asymmetry possibly reaching to higher multipoles \cite{wmap5_evidence}. Otherwise, the nature of the asymmetry and the maximum asymmetry direction remains almost the same, and it will be interesting to see what results from Planck will have to say about an asymmetry at small scales.

These analyses provide motivations for the study of inflationary
models that can generate a spatial asymmetry at low $\ell$ while
remaining isotropic at larger values of $\ell$. If these anomalies
prove to be valid indicators of an asymmetry in the power spectrum,
they can provide a direct probe of inflationary dynamics.
Significant work that attempts to find a solution to these anomalies
has already appeared \cite{Gumrukcuoglu:2006xj}, \cite{models}, \cite{Explantion_north_south},
\cite{other_suppression} in the literature.

 In this paper we discuss a simple situation that could
lead to a spatial asymmetry in the CMB power spectrum at low
values of $\ell$ within single field inflation. This involves an
initial period of fast-roll expansion driven by the inflaton
kinetic energy. The possibility of such an initial fast-roll
period has been proposed by Contaldi, Peloso, Kofman and Linde
(CPKL) \cite{Contaldi:2003zv} as a mechanism to explain the lack
of CMB power at low $\ell$. This mechanism provides a suppression
of the spectrum of primordial perturbations and thus of the CMB
at large scales, and it has also been worked on by others \cite{extrawork}.

We will argue that in situations where the initial kinetic energy
is significant in comparison to the potential energy, we should also
expect the presence of a spatial gradient in the initial conditions
of the inflaton field. We will show that even a surprisingly small
value of an initial gradient -- of order a few percent -- will
leave an observable spatial asymmetry in the CMB power spectrum at
low $\ell$. Essentially, the power suppression in the
fast-roll model occurs at scales that depend sensitively on the
initial magnitude of the scalar field in the frame where the kinetic
energy is uniform and isotropic. This leads to a characteristic
pattern for the spatial dependence of the power spectrum. While we
will provide a brief discussion of two-field models below, we here
focus on the single field fast-roll option because of its simplicity
and predictive power.

\section{Kinetic energy and spatial asymmetries}

Inflation provides an explanation for the isotropy and homogeneity
of the present universe. Rather than having to postulate extremely
smooth initial conditions for the early universe, a long period of
inflation will take non-smooth initial conditions and still lead
to a highly isotropic and homogeneous observable universe today.
However, if the number of e-foldings of inflation is just barely
the minimal number, about 60, the initial conditions could be
relevant and could modify the first few e-foldings.

The CPKL mechanism \cite{Contaldi:2003zv} postulates an initial
period of kinetic energy dominance which then rapidly evolves into
the standard slow-roll paradigm where the potential energy dominates
and the universe inflates.
If the slow-roll phase is many e-foldings longer than the minimum
number of e-folds, the effects of the initial kinetic phase will be
unobservable since the scales associated to its effects will be
stretched far beyond our observable horizon.
However, if the slow-roll phase is close to the minimum number of e-folds, then
that initial kinetic phase will modify the first few e-foldings
that generate the CMB power spectrum on largest scales,
$i.e.$ for small values of $\ell$.

In a universe dominated by a uniform scalar field, the equations of
motion are the Klein-Gordon equation on a FRW background
\begin{equation} \label{KGeq}
 \ddot \phi + 3 H \dot \phi + \frac{d V}{d \phi}(\phi) = 0
\end{equation}
and the Friedman equation\footnote{We use units where $M_{P} = G^{-1/2} = 1$.}
\begin{equation}\label{Friedman}
 H^2 = \left(\frac{\dot a}{a}\right)^2 = \frac{8 \pi}{3} \rho = \frac{8 \pi}{3} \left(\frac{\dot \phi^2}{2}   + V(\phi)\right),
\end{equation}
and the equation of state parameter is
\begin{equation}
 w = \frac{P}{\rho} = \frac{\frac{\dot \phi^2}{2} - V(\phi)}{\frac{\dot \phi^2}{2} + V(\phi)}.
\end{equation}

The CPKL assumption is an initial condition in which the inflaton
velocity, $\dot{\phi}$, is non-zero and the kinetic energy term
is dominant over the potential energy, $\frac{\dot \phi^2}{2} \gg V(\phi)$.
During this initial phase $w \approx 1$ and the expansion of the
universe will be decelerating similar to a matter dominated universe rather
than a deSitter expansion. In this phase the kinetic energy rapidly
decreases, until eventually the potential energy dominates and we
enter the usual slow-roll phase. There is a short transitional
phase when the potential energy already dominates and the universe
inflates, but the inflaton velocity has not yet settled to its slow-roll value
$\dot \phi_{SR} \simeq - \frac{1}{2\sqrt{6 \pi}} \hspace*{1pt} \frac{dV}{d\phi} / V^{1/2}$.

The initial conditions involve specifying both $\phi$ and $\dot{\phi}$
which have to be chosen appropriately to obtain only about 60 e-folds
of inflation so that the effects of the initial fast-roll stage are
observable. CPKL then show that the quantum fluctuations are suppressed
during the onset of inflation when the inflaton is fast-rolling -- this
will be reviewed in the next section. The picture that emerges then
involves suppressed fluctuations at early times followed by standard
slow-roll behavior. Since the earliest times correspond to the largest
scales, the low $\ell$ multipoles are suppressed while the higher ones
are standard.

Our extension of CPKL comes from the observation that the initial
conditions in $\phi$ and $\dot{\phi}$ need not be the same at all
positions in space. If they are close to uniform, one can expand
the values in a multipole expansion. The first deviation from
uniformity would consist of a gradient in the initial conditions
across the initial patch. We will consider only such leading
linear deviations from uniformity in this paper.

CPKL invoke a uniform initial condition in $\dot{\phi}$. Actually
this is not a separate assumption. Because the value of
$\dot{\phi}$ is changing with time, one can always choose a
time-slicing such that $\dot{\phi}$ is uniform across the initial
time slice. That is, if there is a gradient in the initial
condition for $\dot{\phi}$ using one definition of the initial
time slice, one can change to another definition such that this
variable is uniform. However, in this frame there is no a priori
reason for the initial value of the magnitude of $\phi$ itself to
be uniform. A mechanism that can produce a temporal variation in
$\phi$ can in principle also produce a spatial gradient in the
field. We could equally well define a different time frame in
which the initial condition of $\phi$ is uniform, but in this
frame we would in general not expect that the initial value of
$\dot{\phi}$ is constant in space. It is an extra assumption to
assume that the initial conditions for $\phi$ and $\dot{\phi}$ are
spatially uniform in the same frame.

\begin{figure}
\begin{center}
  \begin{minipage}[t]{.07\textwidth}
    \vspace{0pt}
    \centering
    \vspace*{120pt}
    \hspace*{-10pt}
    \rotatebox{90}{$\dot \phi$}
  \end{minipage}%
  \begin{minipage}[t]{0.70\textwidth}
    \vspace{0pt}
    \centering
    \includegraphics[width=0.99\textwidth,height=!]{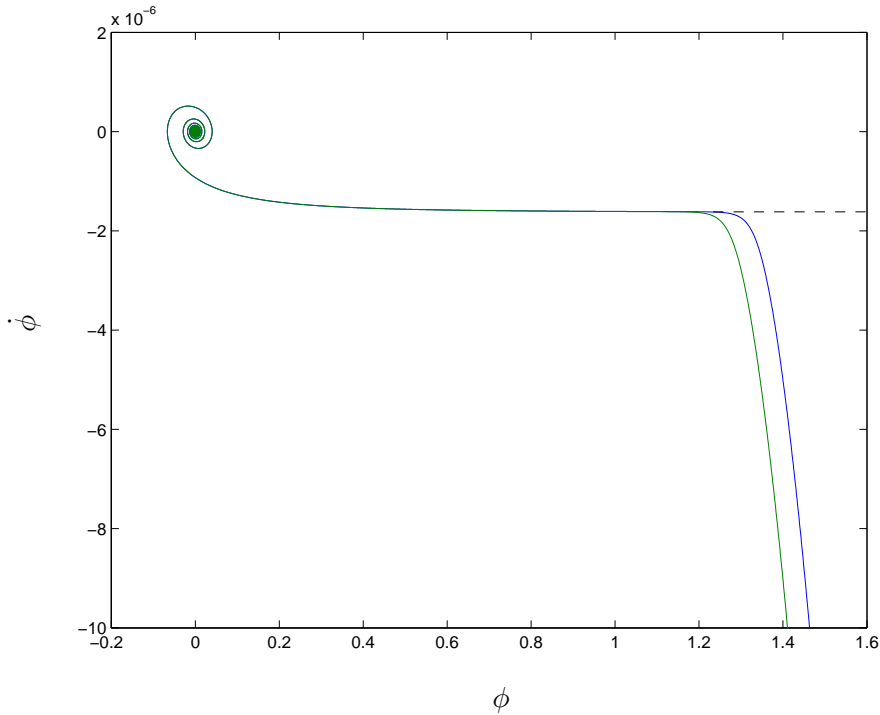}
  \end{minipage}
\end{center}
$\hspace*{240pt} \phi \vspace*{10pt}$ \caption{\small{Phase space
  diagram of inflationary background solutions for two different
  values of the initial scalar field with both initially in the
  kinetic energy domination regime. The dashed curve shows the slow-roll attractor.}}
\label{FigurePhaseSpace}
\end{figure}

We can obtain an invariant description of the evolution of the
inflaton by considering trajectories in the $\phi-\dot{\phi}$
plane, displayed in Fig. \ref{FigurePhaseSpace} for a chaotic
inflation potential $V(\phi) = \frac {1}{2} m^2 \phi^2$. There are
three phases visible in this plot: Initially, kinetic energy
dominates and due to the rapid decrease in the kinetic energy, the
trajectory runs quickly to the slow-roll attractor line where
$\dot{\phi} \simeq \dot \phi_{SR} = const$. All initial
trajectories are attracted to this line - this is one of the key
features of inflation. Finally at the end of inflation, there is
inflaton decay and reheating. However, different starting points
lead to differing amounts of the slow-roll phase -- these are
shown as different initial trajectories. The number of e-folds of
slow-roll inflation is
\begin{equation} \label{NSR}
 N \simeq 2 \pi \phi_{SR,i}^2
\end{equation}
where $\phi_{SR,i}$ is the field value when the phase space
trajectory hits the slow-roll attractor.

If we start in a frame with a
uniform value of $\phi$, which would be represented by a
vertical slice through the initial trajectories, one needs
significantly different initial values of $\dot{\phi}$ in order to
end up at different points on the slow-roll attractor. However, if
we use a frame with uniform initial values of $\dot{\phi}$, which
would be a horizontal slice through the trajectories, only a small
difference in the initial values of $\phi$ are required to produce
distinct trajectories with differing amounts of slow-roll
behavior and thus differing amounts of inflation. We will see that
gradients of order only a few percent are needed for observable effects.
Because the potential energy is subdominant at the initial time,
this is only a very small gradient in the initial energy density.
Therefore, an initial slice with constant $\dot \phi$ is the better
choice since we want to work with a FRW metric that requires a homogeneous
and isotropic energy density.

If the inflaton field is not uniform, the Klein-Gordon equation, Eq.
(\ref{KGeq}), contains an additional term proportional to $\nabla^2
\phi$. For a gradient in the field, $\phi(x) = A + B x$ where $A$
and $B$ are constants, $\nabla^2 \phi = 0$ so that we can neglect
this term and still work with Eq. (\ref{KGeq}). The gradient however
contributes a term to the energy density of the form $(\nabla \phi)^2/2$,
which for a linear gradient in the field $\phi$ amounts to a constant throughout space. There is also an
anisotropic contribution to the pressure proportional to $(\nabla_i \phi)^2$ .
However, the gradient's contribution to the energy density and pressure is
always subdominant for the small amounts of gradients we require
so that we can neglect it. Thus, even in the presence of a small spatial
gradient, the inflaton field evolves independently at each spatial
position. That is, in our approximation the trajectories displayed in Fig. 1 are not
modified by the presence of a spatial gradient.

In our extension with a gradient in the initial conditions, two
points on opposite sides of the universe which started with
different initial values of $\phi$ will have undergone different
amounts of inflation so that the large scale suppression features in
the power spectrum associated to an initial fast-roll stage will
appear at different scales today. A gradient in e-folds of inflation
is the leading effect in our model stretching both the cutoff scale
and space by a different amount in different parts of the universe.
Certainly, it is not strictly correct to use a FRW background metric
and the standard formulas for a uniform and isotropic cosmology.
However, the expansion is uniform and isotropic both during the fast
roll phase when the kinetic energy is dominant, and later during the
slow roll phase. The inhomogeneity only effects the universe during
the very short transition region between fast-roll and slow-roll of
the inflaton. We feel that our treatment captures the leading
effects of an initial gradient without the need to solve exactly the
evolution through the short transition region.

\section{CMB fluctuations at low multipoles}
An initial regime of kinetic energy dominance of the inflaton
before reaching the slow-roll attractor modifies the spectrum
of quantum fluctuations as noted by CPKL. For wavelengths much
smaller than the Hubble scale at the onset of the accelerated
expansion, $H_*^{-1}$, the spectrum is the same as the slow-roll
spectrum because the small wavelength perturbations are
insensitive to the overall background expansion of the Universe at
this time. However, for wavelengths comparable to $H_*^{-1}$ the
spectrum is altered and it exhibits a suppression of power for
larger wavelengths. Intuitively, this suppression can be
understood from the relation $\delta \rho/\rho \sim H^2 /\dot \phi$:
if the inflaton rolls faster initially, the spectrum will be
suppressed at larger scales.

Following CPKL, we use a chaotic inflation model with potential
$V(\phi) = \frac{1}{2}m^2\phi^2$ with $m = 4 \times 10^{-7}$. As
initial conditions, we choose $\phi_{i} = 3.59$ and $\dot \phi_{i} =
-35.9 \hspace*{1.5pt} m$. This particular choice of initial
conditions gives us about $60$ e-folds of inflation. The initial
kinetic energy is 100 times larger than the initial potential energy
so that we start out well in the kinetic energy domination regime.

We use the gauge invariant formalism for cosmological perturbations
\cite{Mukhanov:2005sc, Mukhanov:1990me} and the equations of motion
for the perturbation variables read
\begin{eqnarray}
\delta {\phi_{k}^{''}} + 2 \mathcal H \delta \phi_{k}^{'} +
\left(\frac{k}{a}\right)^2 \delta \phi_{k} + V_{,\phi \phi}
a^2\delta \phi_{k}
&=& 4 \phi_{0}^{'} \Phi_{k}^{'} - 2V_{,\phi}a^2\Phi_{k}  \\
4\pi \phi_{0}^{'}\delta \phi_{k} &=& \Phi_{k}^{'} + \mathcal H \Phi_{k}
\end{eqnarray}
where primes denote differentiation with respect to conformal time
$\eta$ and ${\mathcal H} \equiv a^{'}/a$. We solve these
perturbation equations using numerical mode by mode integration.
As the system is initially in the kinetic energy domination
regime, the vacuum is chosen to be the approximate solution in the
kinetic energy domination stage at an initial time $\eta_i$
\cite{Contaldi:2003zv}
\begin{equation} \label{vac}
v_{k}(\eta_{i}) = \sqrt{\frac{\pi}{8\ \mathcal H_{i}}} \
\sqrt{1 + 2\mathcal H_{i}\eta_{i}} \ H_{0}^{(2)}\left(k\eta_{i} +
\frac{k}{2 \mathcal H_{i}}\right),
\end{equation}
where the Mukhanov-Sasaki variable is
\begin{equation}
v_{k} \equiv a\left(\delta \phi_{k} + \frac{\phi^{'}}{\mathcal
H}\Phi_{k}\right).
\end{equation}
We verified that
a different choice of vacuum, the Hamiltonian diagonalization
vacuum given by \cite{Mukhanov:2005sc}
\begin{equation}
v_{k}(\eta_{i})= \left(k^2 - \frac{z^{''}}{z}\right)^{-1/4}, \ \ \ \
v'_{k}(\eta_{i})= i \left(k^2 - \frac{z^{''}}{z}\right)^{1/4}~~~~~~~\text{with} \
z\equiv \frac{a \phi^{'}}{\mathcal H}
\end{equation}
does not change the resulting power spectrum significantly.

\begin{figure}
\begin{center}
  \begin{minipage}[t]{.07\textwidth}
    \vspace{0pt}
    \centering
    \vspace*{116pt}
    \hspace*{-10pt}
    \rotatebox{90}{$P_{\Phi}(k)$}
  \end{minipage}%
  \begin{minipage}[t]{0.70\textwidth}
    \vspace{0pt}
    \centering
    \includegraphics[width=0.99\textwidth]{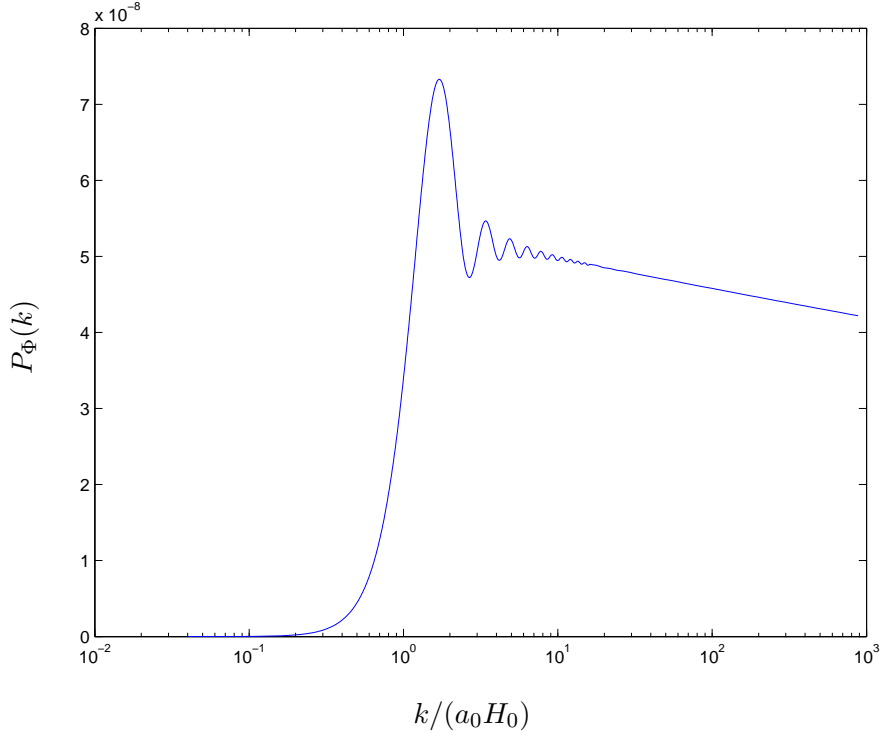}
  \end{minipage}
\end{center}
$\hspace*{210pt} k/(a_{0}H_{0}) \vspace*{10pt}$
\caption{\small{Power spectrum of the gravitational potential
$\Phi$ at the end of inflation in the CPKL model with an initial fast-roll stage.}} \label{FigureSpectrumFastroll}
\end{figure}

In Fig. \ref{FigureSpectrumFastroll} we show the power spectrum of
the gravitational potential after the end of inflation. The cutoff
of the spectrum around $k \sim a_0 H_0$ is the most important
feature of the spectrum, but some oscillations at the transition
to a pure slow-roll spectrum are visible as well. The position of
the cutoff in the spectrum, $i.e.$ the scale associated to the
cutoff, actually depends on the complete expansion history of the
universe since the onset of inflation until today, which is not
exactly known.

Instead of computing the spectrum numerically CPKL have also shown
that an approximate analytic solution with an instantaneous
transition between the regimes of kinetic energy domination and
slow-roll inflation reproduces the exact spectrum extremely well.
This shows that the spectrum does not depend much on the details of
the inflaton potential. We have checked this by generating the
corresponding spectra using $\phi^4$ and $\phi^6$ potentials, and
we found shapes of the resulting power spectra similar to the one
shown in Fig. \ref{FigureSpectrumFastroll} with only small changes
in the width of the cutoff region of the order of 10\%.

Since the main feature of the spectrum is the cutoff that yields suppression
for largest scales, CPKL have also introduced a simple parametrization of a
spectrum with an exponential cutoff
\begin{equation}
 P_{\Phi}(k) \sim k^3|\Phi_{k}|^2 = A_{s}\Big(1 -
 \exp \big[-(k/k_{c})^\alpha\big]\Big) \hspace*{1pt} k^{n_{s}-1} \label{exponentialcutoff}
\end{equation}
with $\alpha = 3.35$ as a useful simplified model.

\begin{figure}
\begin{center}
  \begin{minipage}[t]{.07\textwidth}
    \vspace{0pt}
    \centering
    \vspace*{80pt}
    \hspace*{0pt}
    \rotatebox{90}{$\ell (\ell+1) \, C_\ell$}
  \end{minipage}%
  \begin{minipage}[t]{0.7\textwidth}
    \vspace{0pt}
    \centering
    \includegraphics[width=0.99\textwidth]{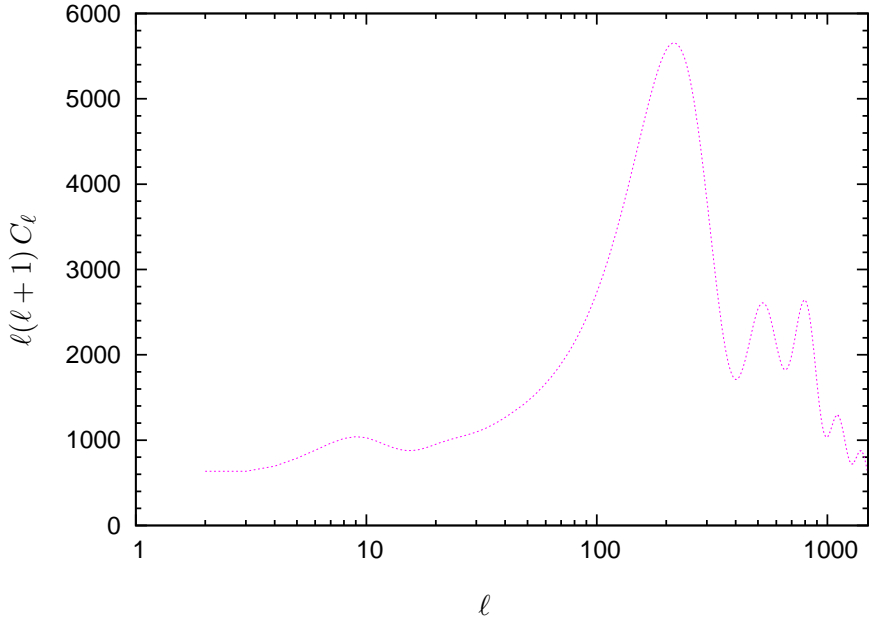}
  \end{minipage}
\end{center}
$\hspace*{240pt} \ell \vspace*{10pt}$ \caption{\small{TT
power spectrum of the CMB in the CPKL model.}} \label{ClSpectrumFastRoll}
\end{figure}

\begin{figure}
\begin{center}
  \begin{minipage}[t]{.07\textwidth}
    \vspace{0pt}
    \centering
    \vspace*{90pt}
    \hspace*{-10pt}
    \rotatebox{90}{$\ell (\ell+1) \, C_\ell$}
  \end{minipage}%
  \begin{minipage}[t]{0.70\textwidth}
    \vspace{0pt}
    \centering
    \includegraphics[width=0.99\textwidth,height=!]{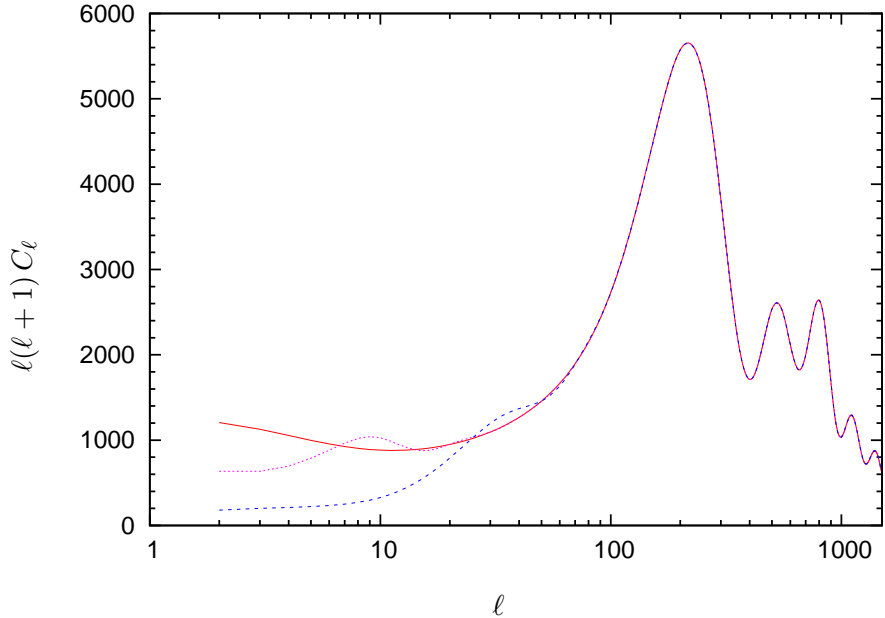}
  \end{minipage}
\end{center}
$\hspace*{240pt} \ell \vspace*{10pt}$ \caption{\small{CMB
spectra for different numbers of e-folds of inflation.
  The solid line comes from $N \gg 60$ so that the cutoff feature
  due to an initial fast-roll stage is stretched beyond our
  observable universe and thus it corresponds to a pure slow-roll spectrum.}}
\label{CltwoSpectraFastRoll}
\end{figure}

The spectrum of CMB fluctuations can be computed from the
primordial power spectrum $P_\Phi(k)$. A variety of astrophysical
processes make this a complicated and highly numerical task which
is usually performed with numerical codes \cite{CMBFAST, CMBEASY, CAMB}.
Since the power spectrum in wavenumber is not modified at large $k$,
the CMB spectrum will be the same as
the standard slow-roll predictions at high multipoles $\ell$.
At low $\ell$ the leading effects are the Sachs Wolfe effect and the late time integrated Sachs Wolfe effect.
For the calculation of the CMB spectrum we have modified the numerical code CAMB \cite{CAMB}
to include the CPKL primordial power spectrum as the initial spectrum.
The resulting CMB spectrum is shown in Fig.
\ref{ClSpectrumFastRoll}.  One of the main results is the fact
that the features in the primordial spectrum get somewhat smoothed
in the CMB spectrum but the shape is generally not altered
very significantly. We see the power suppression at low multipoles due
to the initial fast-roll stage of the inflaton field, and at high multipoles
the spectrum matches the flat slow-roll spectrum. The effect of the wiggles
around the cutoff is also evident in the CMB spectrum. However, we note that unlike
the primordial spectrum of Fig. \ref{FigureSpectrumFastroll}, the CMB spectrum
for multipoles smaller than the scale of the cutoff does not approach
zero as quickly, but seems to have an offset which depends on
the scale of the cutoff, see also Fig. \ref{CltwoSpectraFastRoll}. This offset
arises due to the late time integrated Sachs-Wolfe effect which gives
contributions to $C_\ell$ from a wider range of scales than the Sachs-Wolfe effect does.
In the presence of a cutoff in the spectrum at large scales, the late time
integrated Sachs-Wolfe effect is therefore more important than the Sachs-Wolfe
effect at very low $\ell$ \cite{Finelli:2005zc}.

Since the details of inflation and reheating are not pinned down
precisely, the expansion history of the universe is not known
exactly and the number of e-folds $N$ during inflation enters as an
adjustable parameter, where a minimum of about 60 is needed to solve
the flatness and horizon problems. Changing $N$ in our case will
shift the position of the feature in the spectrum. For $N \gg 60$,
the feature is at scales much larger than our present horizon and
the observable CMB spectrum is indistinguishable from a slow-roll
spectrum. If two different parts of the universe underwent different
amounts of inflation $N$ after the initial fast-roll stage due to
different initial conditions $\phi_i$ and $\dot \phi_i$, their
spectra will match at small scales or large $\ell$ as seen in Fig.
\ref{CltwoSpectraFastRoll}. The power spectrum in the part of the
universe that inflated more e-folds will have the scale associated
to the feature in the spectrum stretched more so that the feature
appears at smaller multipoles $\ell$. There is in addition a
geometric effect due to a temporary asymmetric expansion, which we
explore in the next section.
The CPKL mechanism is reasonably predictive in that there is only
one free parameter, the scale of the feature in the spectrum, which
is determined by the number of e-folds $N$ of inflation and thus by
the relative values of the initial conditions for $\phi$ and
$\dot{\phi}$ -- see Fig. \ref{FigurePhaseSpace} and Eq. (\ref{NSR}).
Observationally, this is manifest as the value of $\ell$ below which
there is a suppression in the CMB power spectrum.

\section{Gradient phenomenology and hemisphere analysis} \label{sec:pheno}

With a spatial gradient in the initial value of $\phi$ in the frame
where $\dot{\phi}$ is uniform, parts with different initial values
undergo different amounts of inflation. Once all portions of
the observable universe are in the slow-roll phase, the standard
inflationary description holds with CMB fluctuations at these scales
and subsequent structure formation being spatially uniform.
The effect of the initial gradient only modifies the transition region
between kinetic domination and slow-roll. This transition yields
the cutoff in the spectrum at the scale $k$ that will be stretched by
different factors in different parts of the sky due to the presence of
the gradient.

Let us consider two different parts of the universe with slightly
different initial field values $\phi_{i,1}$ and
$\phi_{i,2} = \phi_{i,1} + \Delta \phi$. This also gives different
amounts of inflation in the two parts, $N_1$ and $N_2 = N_1 + \Delta N$
where the difference in e-folds scales proportionally to $\Delta \phi$
\begin{equation}
  \Delta N \sim \Delta \phi.
\end{equation}
Normalizing the scale factor at the initial time, a different amount
of e-folds of inflation gives $a_2 / a_1 = e^{\Delta N}$ after inflation
so that the relationship between coordinates and wavenumbers in the
early universe and physical scales today then depends
{\it exponentially} on the difference in e-folds.

Our first task is to understand how a linear gradient in the scalar
field translates to variations in the physical scales on the surface
of last scatter.
Let the coordinate $\zeta$ describe the direction
along which the initial value of the inflaton field has a linear
gradient in the frame where $\dot \phi$ is constant, i.e.
\begin{equation}
 \phi_i (\zeta) = A + B \zeta~.
\end{equation}
A small patch of the initial volume with thickness $d\zeta$ will
inflate to a patch of the sky today with a thickness $dz$ in
physical coordinates, where $z $ measures distance along the same
direction today. Because different patches will have different
amounts of inflation, with $a_2 / a_1 = e^{\Delta N}$ and $\Delta N
\sim \Delta \phi$, the thicknesses will be related by
\begin{equation} \label{eq:dzdzeta}
 dz = \rho e^{b\zeta} d\zeta.
\end{equation}
Here $\rho$ is the rescaling of coordinates due to expansion that
would have happened without the gradient in the scalar field
(with $\phi_i = \phi_i(\zeta=0) = A$): $\rho
\sim a(t_0)/a(t_{i}).$ The parameter $b$ depends on the magnitude of
the gradient $\frac{\Delta \phi}{\Delta \zeta}$ in the inflaton field, with
$b=0$ if there is no gradient in the field. Eq. (\ref{eq:dzdzeta})
shows the geometric effect of the gradient on the expansion of space:
An initial patch located at a higher $\phi_i$ than a second patch and separated
by a distance $\Delta \zeta$ will inflate $\Delta N = b \Delta \zeta$ e-folds
more than the second one, so that the first patch will today be larger in
the $z$-direction by an exponential factor of $e^{\Delta N}$.

The relation in Eq. (\ref{eq:dzdzeta}) can be integrated to relate the
initial coordinates along the gradient to coordinates today,
\begin{equation}
z =\frac{\rho}{b}( e^{b\zeta}-1)
\end{equation}
or
\begin{equation}
b\zeta = \ln \left[1+  \frac{bz}{\rho}  \right]\equiv\ln [1+  b'z ].
\end{equation}
Correspondingly we can relate a feature in the initial wavenumber
spectrum, for example the location of the start of the cutoff in
wavenumber, to the physical scales today. A cutoff that appears at
$k_c$ in the original spectrum would appear at scales
\begin{equation} \label{eq_k_exp}
k= \frac{k_c}{\rho e^{b\zeta}} \equiv k_0 e^{-b\zeta}
\end{equation}
today. Rewriting this in terms of the present physical coordinates
implies that the feature appears at
\begin{equation}
 k = \frac{k_0}{1+b'z}.
\end{equation}
It is apparent that due to the geometric effect of different amounts
of expansion of different parts of the universe an initial gradient
does not yield a simple gradient today! Finally, since we are
interested in the CMB radiation coming from the surface of last
scatter, we are interested in how this feature varies with
direction. If the surface of last scatter is a distance $R_*$ away,
and we use $z= R_* \cos \theta$ and define $a = b' R_*$, we have the
break in the spectrum occurring at
\begin{equation} \label{eq:kCutVar}
 k = \frac{k_0}{1+ a \cos \theta}.
\end{equation}
The result is that an initial gradient in the scalar field, or
equivalently in the start of the slow-roll phase, leads to the
break in the CMB power spectrum occurring at a wavenumber that
depends on the position in the sky by the above relation.

It is tempting to translate Eq. (\ref{eq:kCutVar}) into a relation
for the cutoff position in terms of angles or multipoles in the CMB
spectrum and use it to find the CMB modulation function for this
model with a gradient. This is quite easy to do using the approximate
primordial CPKL spectrum from Eq. (\ref{exponentialcutoff}) and
only taking into account the Sachs-Wolfe effect which results in a
$C_\ell$ spectrum of the same form as Eq. (\ref{exponentialcutoff})
(see the second toy model analysis in the Appendix). However as we
have pointed out in the last section, the late time integrated Sachs-Wolfe
effect is important at low $\ell$, and its inclusion unfortunately does not
seem to result in a simple analytic modulation function.

A more rigorous test of the model would be to calculate the general
correlations $\left<a^*_{\ell' m'} a_{\ell m}\right>$ including
off-diagonal terms which vanish in the homogeneous and isotropic
limit and compare these correlations to CMB data, as proposed for
a different anisotropic model in \cite{Gumrukcuoglu:2006xj}.
Such an analysis would go beyond the scope of this paper. Instead we will
perform a simple first test of our model making use of the existing data
in which the CMB power spectrum is extracted separately from two hemispheres
of the sky.

\begin{figure}
\begin{center}
  \begin{minipage}[t]{.07\textwidth}
    \vspace{0pt}
    \centering
    \vspace*{99pt}
    \hspace*{-10pt}
    \rotatebox{90}{$\ell (\ell+1) \, C_\ell$}
  \end{minipage}%
  \begin{minipage}[t]{0.83\textwidth}
    \vspace{0pt}
    \centering
    \includegraphics[width=0.99\textwidth,height=!]{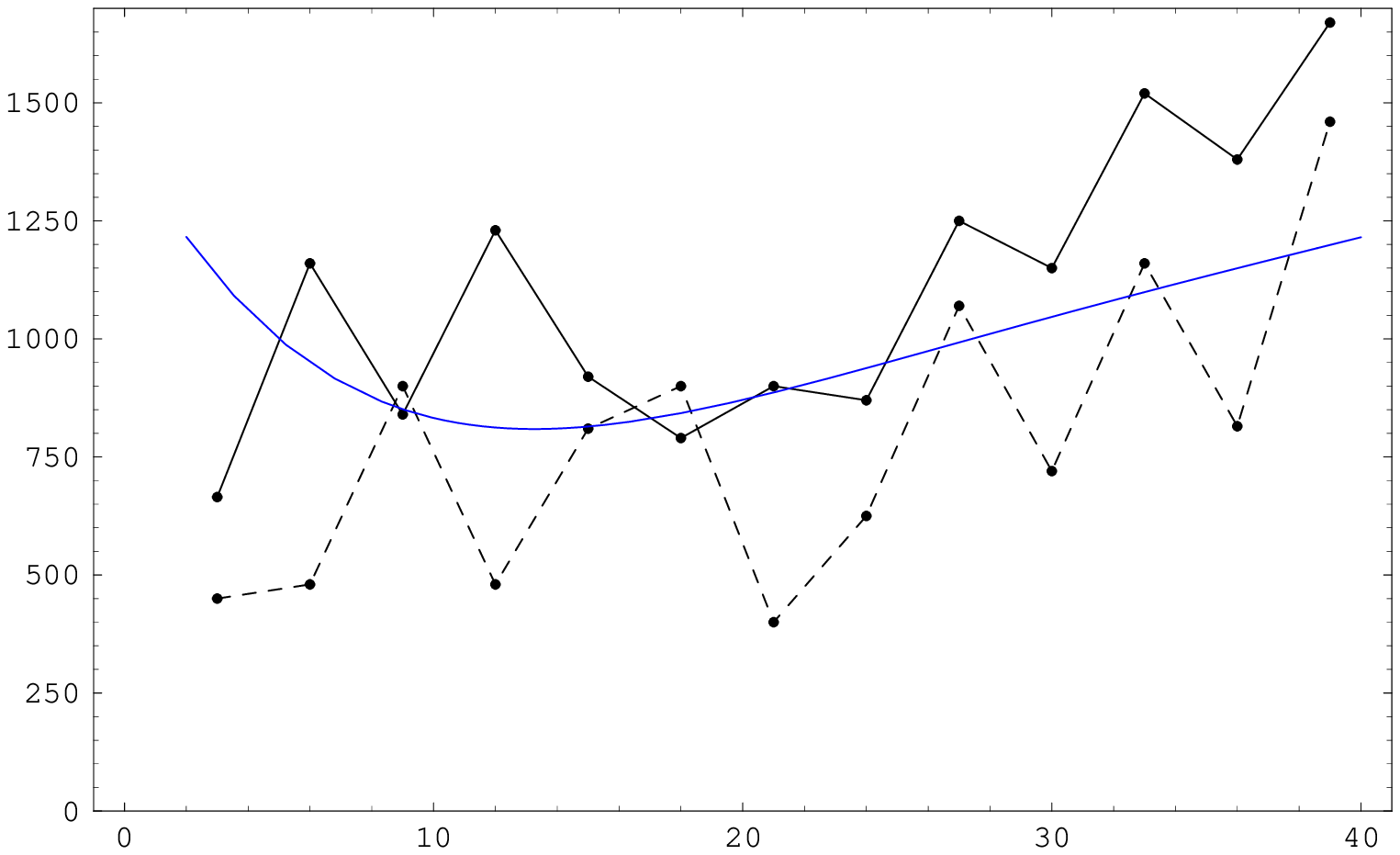}
  \end{minipage}
\end{center}
$\hspace*{240pt} \ell \vspace*{10pt}$ \caption{{\small Sketch of the
measured CMB power spectra (mimicking the right panel of Fig. 7 in
Ref. \cite{Maino:2006pq}) for the northern hemisphere (connected by
dashed lines) and for the southern hemisphere (connected by solid
lines), and the WMAP-3 best fit $\Lambda$CDM spectrum $C_\ell^0$
(smooth blue curve) \cite{Hinshaw:2006ia}.\vspace*{10pt}}}
\label{fig_hemidata}
\end{figure}

Data in hemisphere form has been reported by \cite{Hansen:2004vq,
Hansen:2006rj, Maino:2006pq} and we sketch the results of the
analysis of the WMAP 3-year data of \cite{Maino:2006pq} in Fig.
\ref{fig_hemidata}. The two measured CMB spectra displayed are
extracted from the two hemispheres with respect to the maximum
asymmetry axis pointing to the north pole $(\theta, \phi) =
(80^\circ, 57^\circ)$ in galactic coordinates, and the power
spectrum for $\ell < 40$ obtained from the northern hemisphere
exhibits a lack of power compared to the power spectrum of the
southern hemisphere.

In order to perform a first test of our model with an initial
fast-roll stage and a spatial gradient in the initial field value, we
approximate our results by performing a hemisphere averaging
of the CMB spectrum and compare our predictions to the measured
data in Fig. \ref{fig_hemidata}. For that we orient our initial
gradient in the direction of the maximum asymmetry axis observed.
Furthermore, we identify the point of the gradient with the
lowest field value (where we expect the least amount of inflation
and thus a cutoff feature in the spectrum present at smaller scales
or higher $\ell$ than the rest of the gradient) as the north pole.
That then has the potential to yield a spectrum close to the observed
one with a suppression in the northern hemisphere.
\begin{figure}
 \begin{center}
  \includegraphics[width=0.8\textwidth]{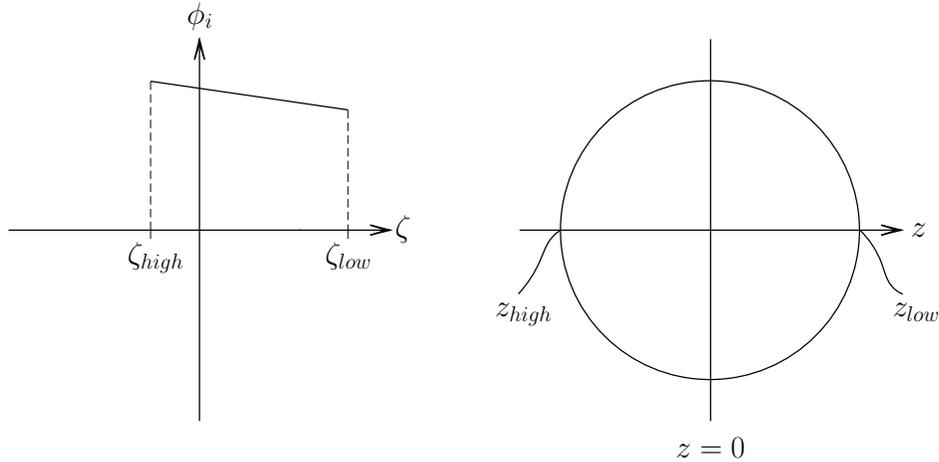}
 \end{center}
 \caption{\small{The gradient in $\phi_i$ shown in the original
coordinates $\zeta$ which are transformed into coordinates $z$ in the
present universe in which the surface of last scattering is shown.}} \label{coordinates}
\end{figure}

Due to the fact that parts of the universe with different initial
field values undergo different amounts of inflation, we first have
to clarify what initial conditions lead to two equally sized
hemispheres today. We start with the northern hemisphere (N) which
contains the point with the lowest initial field value of the
gradient, $i.e.$ the lower part of the gradient, and choose a
certain amount of initial gradient to give the northern hemisphere
today. This setup is sketched in Fig. \ref{coordinates} in the
coordinate $\zeta$ and in today's physical coordinate $z$ which are
aligned with the gradient and the maximum asymmetry axis, and we
normalize the coordinates such that the point $\zeta = 0$ becomes $z
= 0$ today. If at $\zeta = 0$ we expect $N = N_0$ e-folds, at the
north pole, $i.e.$ the lowest point of the gradient $\zeta =
\zeta_{low}$, the amount of inflation is $N = N_0 - \Delta N_{low}$.
Since the number of e-folds scales linearly with the initial field
value, the number of e-folds as a function of the spatial coordinate
$\zeta$ is
\begin{equation} \label{efoldszeta}
 N(\zeta) = N_0 - \Delta N_{low} \hspace*{1pt} \frac{\zeta}{\zeta_{low}}.
\end{equation}
Now we are interested in the size of this hemisphere today which follows
from integrating $dz = e^{\mathcal N} e^{N(\zeta)} d\zeta$ where
$\mathcal N$ is the number of e-folds of the expansion of the universe
since the end of inflation:
\begin{equation}
 z_{low} = e^{\mathcal N} e^{N_0} \zeta_{low} \frac{1-e^{-\Delta N_{low}}}{\Delta N_{low}}
\end{equation}
which gives us the radius of the first hemisphere, i.e. the part of
the gradient that inflated the least. As we want to construct two
hemispheres of equal size, we require the size of the southern
hemisphere (S) which contains the highest part of the gradient
to be equal to the size of the northern hemisphere,
\begin{equation}
 z_{high} = - z_{low}.
\end{equation}
With the relation between the coordinates $\zeta$ and $z$, we
can in turn determine
\begin{equation}
 \zeta_{high} = - \zeta_{low} \hspace*{1pt} \frac{\log 2 - e^{- \Delta N_{low}}}{\Delta N_{low}}
\end{equation}
and find the number of e-folds for the point with most
inflation from Eq. (\ref{efoldszeta}) to be
\begin{equation}
 N(\zeta_{high}) = N_0 + \log \left(2 - e^{- \Delta N_{low}}\right).
\end{equation}
This result is interesting because it means that no matter how
large a gradient there is in the northern hemisphere, the gradient in the
southern hemisphere will at most give a difference of
$\log 2 \approx 0.69$ e-folds of inflation within the southern hemisphere.
This illustrates that parts of the universe with lower initial field values
expand exponentially less than parts with higher field values so that
the parts with highest initial field values dominate the universe today
and parts with lowest initial field values comprise only a tiny fraction of
the universe.

For a power spectrum with a feature that varies as a function of
position $\zeta$ or $z$, we have to average the power spectrum over
the two hemispheres in order to compare with the measured spectrum
sketched in Fig.\ref{fig_hemidata}. Our approximation to this
averaging is to imagine to cut each hemisphere into slices. If we use
slices of equal thickness $\Delta z$ in today's coordinates, we take the
average over all slices of the hemisphere by weighing
each slice by its number of measured points on the surface
of last scattering. The number of measured points is assumed to be
proportional to the surface area of a slice, and since the surface
area of a spherical segment only depends on its thickness $\Delta z$ and
the radius of the sphere, each slice of equal thickness has the same
weight in the average over the hemisphere.

In the Appendix we present a simple toy model with a step function cutoff
as the primordial fluctuation spectrum. It is illustrative since most
parts of the calculation can be performed analytically.
Our analysis here however is performed completely numerically. 
We approximate our gradient by a series of equidistant steps and calculate the $C_\ell$ spectra for each of the steps with a constant $\phi_i$, $i.e.$ we use equidistant (in coordinate $\zeta$) slices of the gradient in which we approximate the initial field as constant. When averaging over a hemisphere, we have to average over all steps contained in the hemisphere. The geometric effect
of different amounts of expansion of the different slices is accounted
for by the weight factor
$\exp\left(-\Delta N_{low} \, \frac{\zeta}{\zeta_{low}}\right)$
for each step, where we take the position $\zeta$ of the slice as
its center value. 
The effect of our mechanism on the averaged CMB spectra of the two hemispheres
is best illustrated by defining relative suppression functions
\begin{equation}\label{defF}
 F_\ell^D = C_\ell^D / C_\ell^{SR}
\end{equation}
with $D=N,S$ as the ratio of the CMB spectra for the two
hemispheres and the CAMB spectrum we obtain for pure
slow-roll inflation.

The two free parameters then in this hemisphere averaging procedure
are the initial field values $\phi_i(\zeta=0)$ and $\phi_i(\zeta=\zeta_{low})$,
which then determine $\phi_i(\zeta=\zeta_{high})$ as well as $N_0$, $\Delta N_{low}$ and $\Delta N_{high}$.
In order to find a good fit to the hemisphere data in Fig. \ref{fig_hemidata},
we first generate a large array of slice spectra using CAMB. Here, we use slices
which have a thickness $\Delta \zeta$ corresponding to $\Delta N_{slice} = 0.1$,
$i.e.$ the difference in number of e-folds between two adjacent slices is $0.1$.
In order to achieve the best fit to the data points for the two hemispheres,
we vary the number of slices for the hemispheres (mostly the number of slices
for the northern hemisphere since $\Delta N_{high} \approx 0.7$ for almost
all of parameter space) and we vary the location of the border between
the two hemispheres within our array of slices. That then corresponds to varying $\phi_i(\zeta=0)$ and
$\phi_i(\zeta=\zeta_{low})$. For each possibility, we obtain the suppression
functions $F_\ell^N$ and $F_\ell^S$ for both hemispheres. These are then multiplied
by the best fit $\Lambda$CDM curve $C_{\ell}^0$ for the WMAP 3-year data
(see the smooth curve in Fig. \ref{fig_hemidata}) to yield our
mechanism's prediction for each hemisphere
\begin{equation} \label{eq_theory_Cl_hemis}
 C_\ell^{Theory, \, D} = F_\ell^D \, C_{\ell}^0
\end{equation}
with $D = N,S$. Here, $C_\ell^{Theory, \, D} = C_\ell^D$
if all parameters in CAMB are selected exactly as
for the WMAP 3-year best fit. Since we use the slow-roll
inflation spectrum in CAMB to obtain $C_\ell^{SR}$ instead of a parameterized
spectrum with a different spectral index, there could be small deviations due
to the normalization of the CMB spectra, and we correct for these
by using Eq. (\ref{eq_theory_Cl_hemis}). The optimal
configuration is then found from a $\chi^2$ fit to the data points
in Fig. \ref{fig_hemidata} calculating
\begin{equation} \label{chisquare}
 \chi^2 = \sum_{D=N,S} \sum_\ell \frac{\left(C_\ell^{Theory,D}-C_\ell^{Data,D}\right)^2}{\left(\sigma^D_\ell\right)^2},
\end{equation}
where the variance is taken as the cosmic variance of the
theoretical spectrum of the hemisphere, $\sigma^D_\ell = \sqrt{2/(2\ell+1)} \,
C_\ell^{Theory,D}$.

\begin{figure}
\begin{center}
  \begin{minipage}[t]{.07\textwidth}
    \vspace{0pt}
    \centering
    \vspace*{99pt}
    \hspace*{-10pt}
    \rotatebox{90}{$F_\ell^D$}
  \end{minipage}%
  \begin{minipage}[t]{0.83\textwidth}
    \vspace{0pt}
    \centering
    \includegraphics[width=0.99\textwidth,height=!]{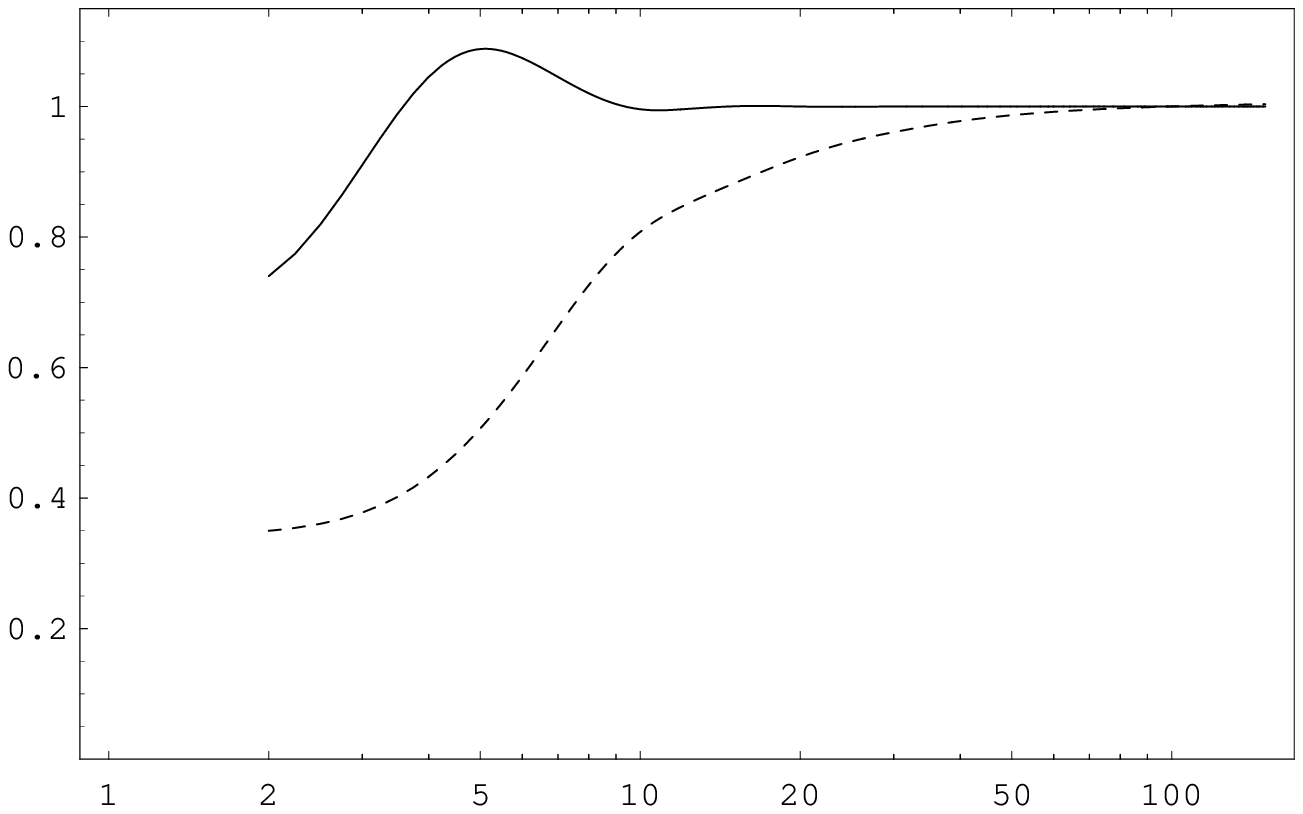}
  \end{minipage}
\end{center}
$\hspace*{240pt} \ell \vspace*{10pt}$
\caption{{\small Suppression functions resulting from averaged CPKL spectra for the northern (dashed) and southern (solid) hemisphere for
         $\Delta N_{low} = 3.6$.}\vspace*{10pt}} \label{fig_NS_full_sup}
\end{figure}

The best-fit configuration has $\Delta N_{low} = 3.6$ and $\Delta N_{high} = 0.7$
which corresponds to less than $4\%$ of variation in $\phi_i$ over both hemispheres
together. In Fig. \ref{fig_NS_full_sup} we display the resulting suppression
functions $F_\ell^D$ for both hemispheres. The spectrum
of the southern hemisphere for the best fit is close to a slow-roll
spectrum with little suppression only at very largest scales.
For the northern hemisphere, the spectrum shows a suppression which
is most significant at $\ell < 10$. Due to the large range
of e-folds $\Delta N_{low} = 3.6$, the averaged spectrum of the northern
hemisphere does not exhibit any peak because the peaks of the
individual spectra of the slices average out.

When compared to the uniform and isotropic $\Lambda$CDM slow-roll CMB spectrum, we
find that $\chi^2$ decreases, $i.e.$ improves, by 3.1 for our model
in the optimal configuration. Moreover, the prediction for the quadrupole
$\ell = 2$ for the full sky sphere is reduced by $45 \%$ in our
model compared to the $\Lambda$CDM best fit prediction. However,
this does not indicate that a significant preference for our
fast-roll with gradient model arises: even though the decrease of
$\Delta \chi^2 = -3.1$ improves the $\chi^2$, the model introduces
four new parameters when compared to the $\Lambda$CDM model --
two angles for the direction of the gradient, the amount of
gradient used $\Delta N_{low}$ and a scale $k_0$ at which the
cutoff appears in a uniform spectrum of one of the slices.

As one can see from Fig. \ref{fig_NS_full_sup}, a significant
suppression only arises for the lowest mutipoles $\ell \mathrel{\rlap{\lower4pt\hbox{\hskip1pt$\sim$}}
\raise1pt\hbox{$<$}} 10-20$. The reason for that is the geometric effect
of the gradient which causes the regions of the gradient with higher
initial field values to dominate the universe, so that the resulting
averaged hemisphere spectra still exhibit relatively steep cutoffs
which are relatively close to each other. For the suppression
functions of the hemispheres for the best fit shown in Fig. \ref{fig_NS_full_sup}
the cutoff in the northern hemisphere's spectrum is roughly at $\ell \sim 2$
whereas for the southern hemisphere it is at $\ell \sim 7$.
That makes it hard to improve the $\chi^2$ more significantly, since
the hemisphere data sketched in Fig. \ref{fig_hemidata} can be
roughly characterized as a constant suppression of order $20 \%$
up to $\ell \sim 40$. Moreover, the errors dominated by cosmic variance
are reduced for suppressed theoretical spectra so that scattered data
points can easily spoil the $\chi^2$.

In the Appendix we additionally discuss two toy models. The first one
uses a step-function primordial spectrum, which allows us to perform
parts of the hemisphere averaging analytically without having to approximate
the gradient as steps. As a second toy model we consider a primordial
spectrum of the same form as Eq. (\ref{exponentialcutoff})
but with a significantly softer cutoff than the CKLP spectrum. Such a
more gradual cutoff could potentially achieve an asymmetry reaching
to higher values of $\ell$.

\section{Conclusions}

The data appears to suggest a spatial modulation in the CMB power
spectrum. However, this asymmetry is unusual in that it may only be present
at largest scales or multipoles
$\ell \mathrel{\rlap{\lower4pt\hbox{\hskip1pt$\sim$}} \raise1pt\hbox{$<$}} 40$.
If inflation lasts only about 60 e-folds, this can occur
if at the earliest stages of inflation there is an asymmetry which however
disappears at later times of inflation. Provided the later stages of inflation
are governed by a single slowly rolling field, the power spectrum at
high $\ell$ and the universe today will be homogeneous and isotropic.
However the quantum fluctuations at the earliest times, and hence the
power spectrum at low multipoles, will show evidence of the initial lack
of isotropy.

We have provided a model of how this situation could have developed
within the framework of single field inflation. It is based on the
CPKL mechanism for large scale power suppression due to an initial
fast-roll stage paired with a gradient in the initial values of the
inflaton field. A gradient in the initial conditions is manifest as differing
numbers of e-foldings of slow-roll behavior on different sides of the universe.
This amounts to shifts in the $\ell$ values at which the suppression
of the CMB power at large angular scales occurs. This model has a distinctive
pattern for the generation of spatial asymmetries which makes it very predictive,
and apart from the asymmetry axis, there are only two additional parameters
characterizing the position of the cutoff and the amount of gradient
within the observable universe.

Our simple analysis where we approximated the initial gradient by a series
of steps and averaged these over two hemispheres provided a first comparison
to WMAP data extracted from different hemispheres. While our model
reproduces the qualitative features of the data and improves the
fit by $\Delta \chi^2 = -3.1$, we cannot claim a preference for our model
since it involves the introduction of in total four new parameters. Because
of the geometric effect on the expansion of space resulting from differing
amounts of inflation due to the initial gradient, a much better fit in a
hemisphere analysis is hard to achieve with our simple single field model with an
initial fast-roll stage and a gradient. Nevertheless, one could improve
on our analysis by performing  a full covariance analysis of the model including all anisotropic degrees of freedom, such as studied in \cite{Gumrukcuoglu:2006xj}. In this way, one would use the
information contained in the general correlations $\left<a^*_{\ell' m'} a_{\ell m}\right>$,
including off-diagonal terms which vanish in the homogeneous and isotropic
limit.

It is also possible to generate spatial asymmetries in the power
spectrum through the use of two or more fields. In this case, it
is possible to assume that one field has a spatial gradient in the
frame in which the other field is uniform. Here there is a lot
more freedom. With two fields, each field has its own potential
with the possibility of a cross coupling between the fields. These
possibilities enable one to modify the shape of the inflaton
potential in different ways either enhancing or suppressing the
rolling of the inflaton field and also modifying the amount of
quantum fluctuations. The resulting possibilities for generating
asymmetries deserve further study. Such models are more flexible,
but are inherently less predictive than the single field model
studied in this paper. However, at least some of these theories
could have a similar phenomenology. By providing an initial faster
evolution, one can suppress the curvature fluctuations providing a
cutoff in the fluctuation spectrum, and a gradient in this initial
condition would survive for only a few e-foldings. With the additional
freedom in multi field models, one may be able to find a model
where the geometric effect which limits our single field
model's capability to obtain a significant asymmetry between
hemispheres extending beyond $\ell \sim 10-20$ is absent or less
restrictive.

We note also that our model, and possible generalizations, have the
potential to impact the analyses of other anomalies. For example,
the cold spot uncovered in \cite{coldspot} occurs along the same
axis as the power spectrum modulation. In our model the primary
effect is the lack of power in some direction. Our hemisphere fit
suggested that one hemisphere is close to the pure slow-roll while
the other hemisphere shows the suppression. This could be similar
in effect to the existence of a cold spot. Likewise, it is
possible that this mechanism can modify the analysis of the
unusual quadrupole-octopole alignments. When attempting a partial
wave decomposition, an overall dipole shifts the underlying power
from one multipole to neighboring ones with $\Delta \ell = \pm 1$.
Our modulation in the spectrum is not exactly a dipole, but could
nevertheless shift the apparent power from one multipole to nearby
ones.

It would be interesting to see if the addition of this form of
spatial gradient to the studies that fit the CMB spectrum can
provide further understanding of the proposed anomalies that are
discussed in the literature. For that a more involved comparison
to data than ours is needed. If the mechanism was successful
it would add to our understanding of inflation and would
increase the confidence in the inflationary paradigm.

\vspace{0.9cm} \centerline{\bf Acknowledgments} \vspace{0.3cm} We
sincerely thank Lorenzo Sorbo and Marco Peloso for useful
discussions, and an anonymous referee for useful suggestions
which resulted in an improved numerical analysis. This work has
been supported partly by the U.S National Science Foundation under
grant PHY-0555304. K.D acknowledges partial support by the DFG cluster of excellence (Munich)
``Origin and Structure of the Universe''. AR has been supported in part by the Department of Energy grant No. DE-FG-02-92ER40704.

\section*{Appendix}

In the Appendix, we study two toy models. The one considered
first with a step function cutoff spectrum is instructive
since the hemisphere averaging for the primordial power spectrum
can be obtained analytically. For the second toy model we use an
exponential cutoff spectrum which falls off more gradually than the
CPKL spectrum.

We construct the theta function toy model keeping the main aspects
of the gradient fast-roll model, a large scale cutoff in the power
spectrum whose scale varies within the universe due to a gradient
in the number of e-folds of inflation. Thus, the scale of the
cutoff in the spectrum varies spatially according to Eq.
(\ref{eq_k_exp}), but we will approximate the spectrum of the
uniform fast-roll model of Fig. \ref{FigureSpectrumFastroll} by a
simple theta function cutoff spectrum $P_{\Phi}(k) = \Theta (k -
k_c)$. For simplicity, we do not include a spectral index and we
compare the resulting spectra to a scale invariant Harrison-Zel'dovich
spectrum.

\begin{figure}
\begin{center}
  \begin{minipage}[t]{.07\textwidth}
    \vspace{0pt}
    \centering
    \vspace*{89pt}
    \hspace*{-10pt}
    \rotatebox{90}{$\bar P^N_\Phi(k)$}
  \end{minipage}%
  \begin{minipage}[t]{0.83\textwidth}
    \vspace{0pt}
    \centering
    \includegraphics[width=0.99\textwidth,height=!]{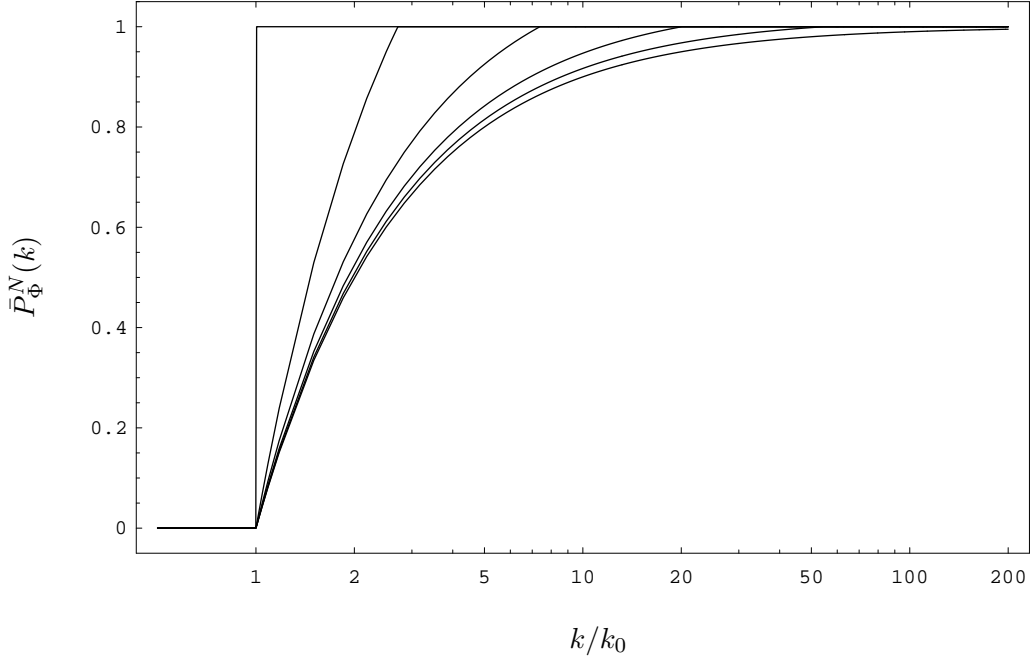}
  \end{minipage}
\end{center}
$\hspace*{240pt} k/k_0 \vspace*{10pt}$ \caption{{\small Averaged theta
function spectra for the northern hemisphere for differing amounts
of gradients such that $\Delta N_{low} = 0,1,2,3,4,\infty$.} \vspace*{10pt}} \label{fig_N_theta_av}
\end{figure}

In the hemisphere analysis in Sec. \ref{sec:pheno} we first calculated
CMB spectra and then performed an average over the two hemispheres. Here
however, we will first average the primordial spectrum
over the hemispheres and then calculate the CMB power spectra. The
primordial spectra for the northern and southern
hemispheres are
\begin{align} \label{eq_av_spec_hemis}
 \bar P^N_{\Phi}(k) & = \int_0^{z_{low}} dz \, P_{\Phi}(k,z) = \int_0^{\zeta_{low}} d\zeta \, \exp\left(-\Delta N_{low} \, \frac{\zeta}{\zeta_{low}}\right) \, P_{\Phi}(k,\zeta) \nonumber \\
 \bar P^S_{\Phi}(k) & = \int_{z_{high}}^0 dz \, P_{\Phi}(k,z) = \int_{\zeta_{high}}^0 d\zeta \, \exp\left(-\Delta N_{low} \, \frac{\zeta}{\zeta_{low}}\right) \, P_{\Phi}(k,\zeta)
\end{align}
from which we can calculate the averaged CMB spectra $C_\ell^N$
and $C_\ell^S$.
As seen in Eq. (\ref{eq_k_exp}) today's physical scale of the
cutoff in the spectrum varies exponentially such that the spectrum becomes
\begin{equation} \label{eq_Thetaspectrum_zeta}
 P_{\Phi}(k, \zeta) = \Theta \left[k - k_0 \exp\left(+\Delta N_{low} \, \frac{\zeta}{\zeta_{low}}\right)\right]
\end{equation}
where the position of the cutoff in the northern hemisphere varies
from $k_0$ to $k_{low} = k_0 e^{\Delta N_{low}}$ depending on the
position $\zeta$. Plugging this into Eq. (\ref{eq_av_spec_hemis}) we
obtain an average spectrum
\begin{equation}
 \bar P^N_{\Phi}(k) = \frac{1-k_0/k}{1-k_0/k_{low}} \ \Theta(k-k_0) \, \Theta(k_{low}-k) + \Theta(k-k_{low})
\end{equation}
for the northern hemisphere which vanishes for $k<k_0$ and equals
unity for $k>k_{low}$.

In Fig. \ref{fig_N_theta_av} we display the averaged spectrum of the
northern hemisphere for different amounts of gradient, $i.e.$ for
different $\Delta N_{low}$, where the limit $\Delta N_{low} = 0$
corresponds to no gradient at all and we recover a pure theta
function spectrum, whereas the limit $\Delta N_{low} = \infty$
corresponds an infinite amount of gradient within the hemisphere. As
one would expect, the average spectrum of the hemisphere is
dominated by the part of the gradient with most e-folds of inflation
which will dominate and make the average power spectrum rather steep
around $k_0$. At $k/k_0  = 10$ the spectrum already reaches $0.9$ in
the limit of an infinite gradient, and since $\ell \sim k$ in the
CMB spectrum, we expect that the cutoff in the resulting averaged
$C_\ell$ spectrum is also rather steep with a range of significant
suppression up to roughly $\ell \sim 10$.

\begin{figure}
\begin{center}
  \begin{minipage}[t]{.07\textwidth}
    \vspace{0pt}
    \centering
    \vspace*{89pt}
    \hspace*{-10pt}
    \rotatebox{90}{$\bar P^D_\Phi(k)$}
  \end{minipage}%
  \begin{minipage}[t]{0.83\textwidth}
    \vspace{0pt}
    \centering
    \includegraphics[width=0.99\textwidth,height=!]{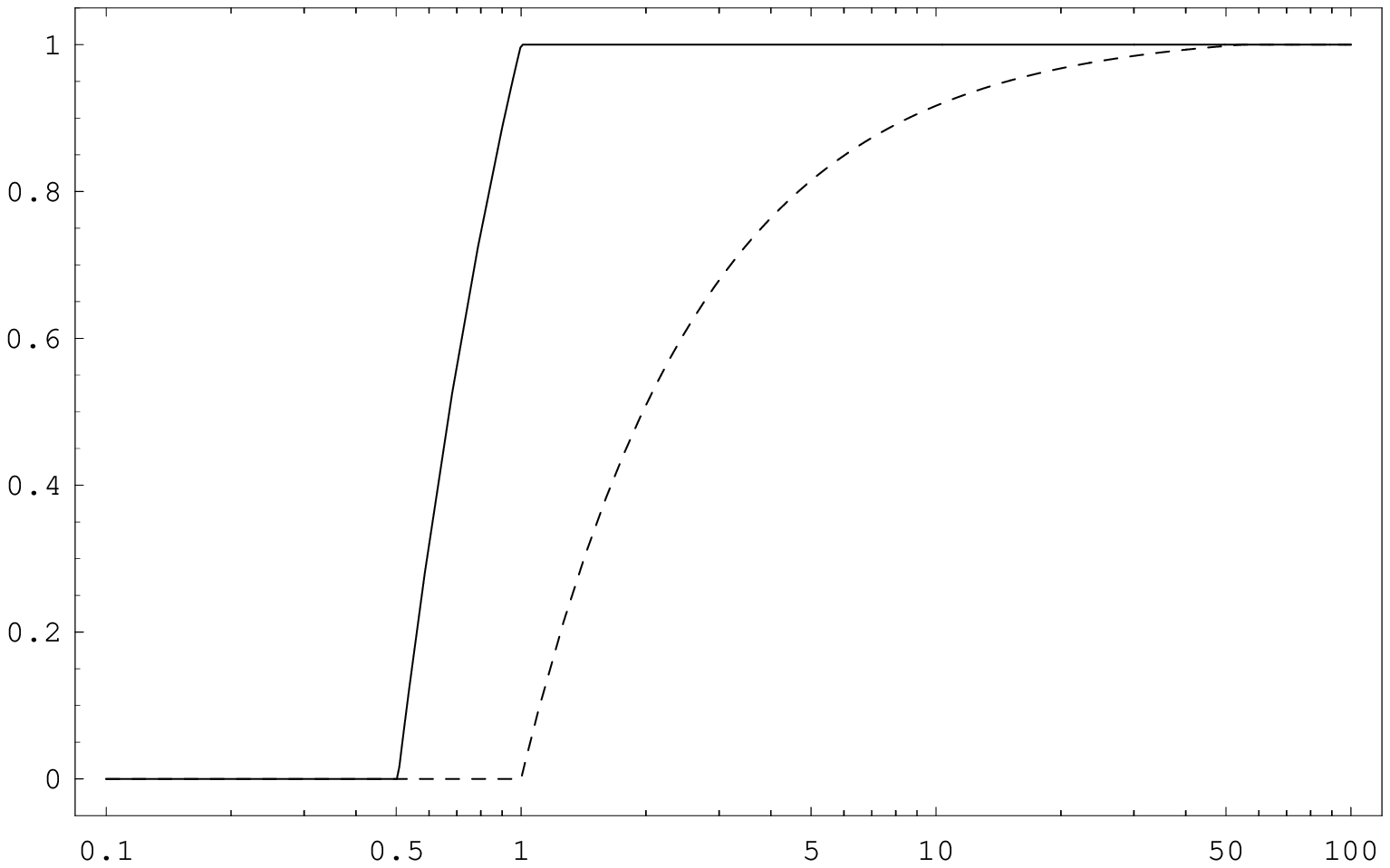}
  \end{minipage}
\end{center}
$\hspace*{240pt} k/k_0 \vspace*{10pt}$ \caption{{\small Averaged theta
function spectra for the northern (dashed) and southern (solid)
hemisphere for $\Delta N_{low} = 4$.}\vspace*{10pt}}
\label{fig_NS_theta_av}
\end{figure}

The averaged spectrum for the southern hemisphere can be calculated
analogously, and for $\Delta N_{low} = 4$, we show both averaged
spectra in Fig. \ref{fig_NS_theta_av}. The averaged spectrum of the
southern hemisphere is steeper than the spectrum of the
northern hemisphere since it contains a much smaller piece of the
initial gradient with a difference of only about $0.69$ e-folds
inside the southern hemisphere as compared to $\Delta N_{low} = 4$
e-folds of difference in the northern hemisphere.

\begin{figure}
\begin{center}
  \begin{minipage}[t]{.07\textwidth}
    \vspace{0pt}
    \centering
    \vspace*{99pt}
    \hspace*{-10pt}
    \rotatebox{90}{$F_\ell^D$}
  \end{minipage}%
  \begin{minipage}[t]{0.83\textwidth}
    \vspace{0pt}
    \centering
    \includegraphics[width=0.99\textwidth,height=!]{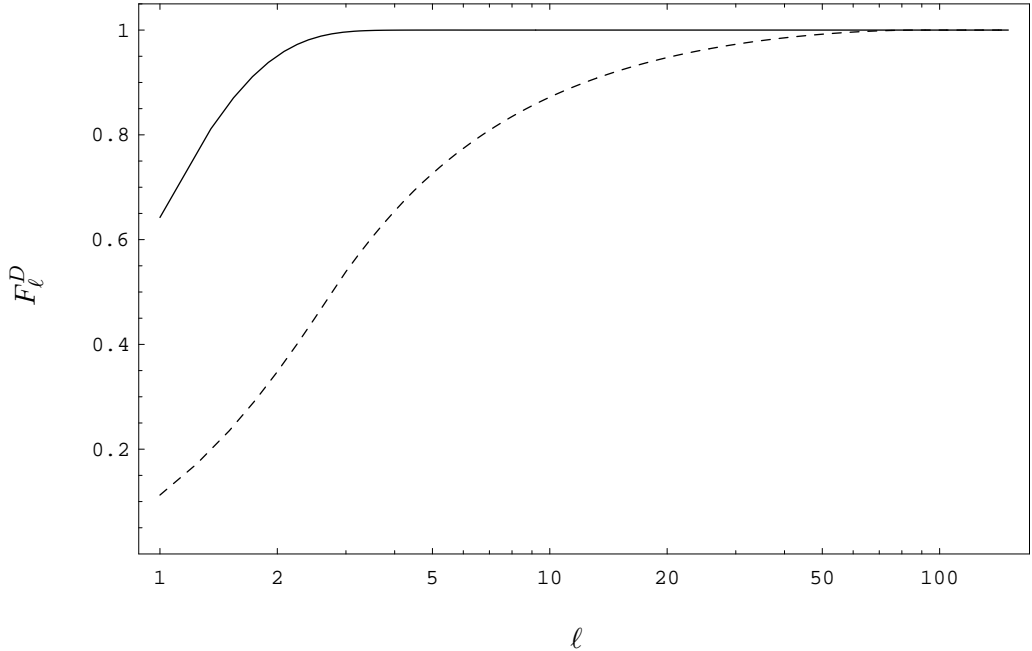}
  \end{minipage}
\end{center}
$\hspace*{240pt} \ell \vspace*{10pt}$ \caption{{\small Suppression functions
resulting from an averaged theta function spectra for the northern
(dashed) and southern (solid) hemisphere for $\Delta N_{low} =
4$. Note that the underlying calculation neglected the integrated Sachs-Wolfe effect. \vspace*{10pt}}} \label{fig_NS_theta_sup}
\end{figure}

As the next step we numerically calculate the CMB spectrum $C_\ell$,
where for this toy model we will only account for the Sachs Wolfe effect \cite{Sachs:1967er} \cite{LiddleandLyth}
\begin{equation}\label{eq_sachs_wolfe}
 C_{\ell} =\frac{4\pi}{9}\int_{0}^{\infty}\frac{dk}{k} \ j_{\ell}^2\left(\frac{2k}{H_{0}}\right)P_{\Phi}(k),
\end{equation}
ignoring all other effects such as the integrated Sachs-Wolfe effect.
These Sachs-Wolfe spectra for the northern and the southern hemispheres
are divided by the Sachs-Wolfe spectrum of a scale invariant Harrison Zel'dovich
spectrum to find the suppression functions $F_\ell^N$ and $F_\ell^S$
for both hemispheres. For that we have to specify the scale $k_0$.
We choose $k_0 = a_0 H_0$ such that the CMB spectrum of the northern
hemisphere ($z=0\dots z_{low}$) will exhibit a suppression of power
at largest scales, whereas the CMB spectrum of the southern
hemisphere $z=z_{high}\dots0$ will exhibit (almost) no large scale
suppression. The resulting suppression functions  $F_\ell^N$ and
$F_\ell^S$ are shown in Fig. \ref{fig_NS_theta_sup}, and as one
would expect, their shapes do not differ much from the primordial
averaged power spectra in Fig. \ref{fig_NS_theta_av}. At small
scales or high $\ell$, both spectra match onto the CMB spectrum from
a scale invariant primordial spectrum, and for observable multipoles
$\ell \ge 2$ the southern spectrum doesn't significantly vary from a
featureless spectrum. The spectrum of the northern hemisphere
however is significantly suppressed at low $\ell$, but as expected,
for $\ell \mathrel{\rlap{\lower4pt\hbox{\hskip1pt$\sim$}}
\raise1pt\hbox{$>$}} 12$ the suppression becomes smaller than $10
\%$.

We can test our theta function cutoff toy model by calculating the
$\chi^2$ of the measured CMB data points in Fig. \ref{fig_hemidata}
with respect to the averaged spectra of the two hemispheres.
The resulting $\chi^2$ for the northern hemisphere decreases by $2.6$ in
comparison to the isotropic $\Lambda$CDM model whereas $\chi^2$ doesn't change
significantly for the southern hemisphere. Unlike in Sec. \ref{sec:pheno}
where we fitted the parameters and obtained $\Delta \chi^2 = - 3.1$ for the
best fit, here we simply chose plausible parameters $k_0$ and $\Delta N_{low}$.

For this toy model, we have only considered the Sachs-Wolfe contributions
to the CMB anisotropies, ignoring the integrated Sachs-Wolfe effect etc.
Nevertheless, when we compare the hemisphere suppression functions in Figs.
\ref{fig_NS_full_sup} and \ref{fig_NS_theta_sup}, we note that
neither the precise form of the cutoff nor the inclusion of the
integrated Sachs-Wolfe effect seem to change the main features.
At the very lowest multipoles $\ell < 5$ one can see in Fig. \ref{fig_NS_full_sup}
the effect of the integrated Sachs-Wolfe effect resulting in a plateau
instead of a cutoff extending down to zero. The main features however
are determined by the geometric effect due to a gradient in the number
of e-folds and by the steepness the cutoff feature in the spectrum.

\begin{figure}
\begin{center}
  \begin{minipage}[t]{.07\textwidth}
    \vspace{0pt}
    \centering
    \vspace*{115pt}
    \hspace*{-10pt}
    \rotatebox{90}{$F_\ell$}
  \end{minipage}%
  \begin{minipage}[t]{0.83\textwidth}
    \vspace{0pt}
    \centering
    \includegraphics[width=0.99\textwidth,height=!]{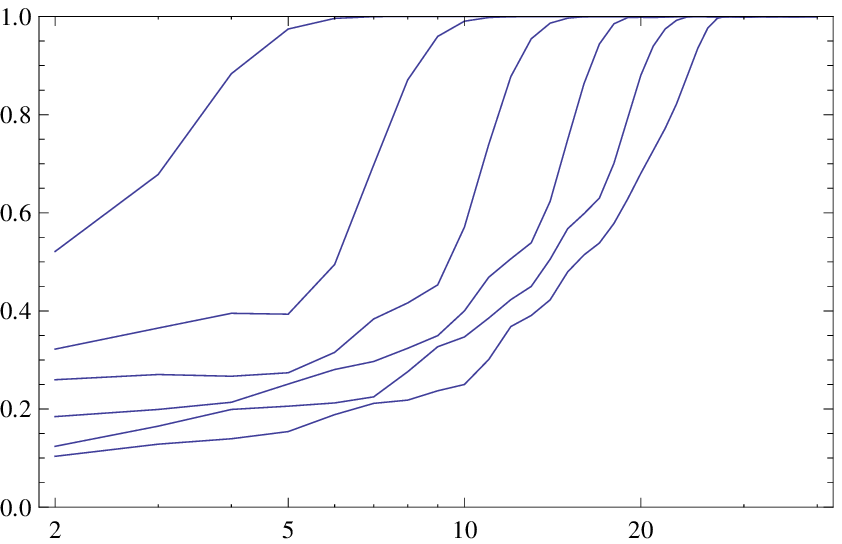}
  \end{minipage}
\end{center}
$\hspace*{240pt} \ell \vspace*{10pt}$ \caption{{\small Suppression functions from fully numerical CMB spectra
resulting from {\it uniform} primordial theta function spectra with the cutoff position varied linearly in $k$.
The plateau to the left of each cutoff arises due to the integrated Sachs-Wolfe effect,
and its height depends on the location of the cutoff.\vspace*{10pt}}} \label{fig_NS_theta_isw}
\end{figure}

Before proceeding to study the next toy model, we will briefly
comment on the integrated Sachs-Wolfe effect which we omitted in the
analysis of the above toy model. As mentioned earlier, the integrated
Sachs-Wolfe effect gives contributions to $C_\ell$ from a wider range
of scales than the Sachs-Wolfe effect does. Therefore at scales larger
than the scale associated with the cutoff, the Sachs-Wolfe effect contributes
almost nothing to the CMB power spectrum but the integrated Sachs-Wolfe effect
can still yield some power in that region. That gives the plateaus or offsets
in the CMB spectrum. To illustrate these plateaus, it is most transparent
to show the CMB spectrum resulting from a uniform theta function primordial
spectrum, as seen in Fig. \ref{fig_NS_theta_isw} for various cutoff
scales. We observe that the height of the offset depends on the scale of the
cutoff where smaller cutoff scales result in a lower offset.

At the other extreme from a step function cutoff is one for which
the cutoff can be made gradual. This can be modeled by use of the
exponential cutoff spectrum introduced in Eq.
(\ref{exponentialcutoff}) where we allow for the exponent $\alpha$
to take on different values. The parameter $\alpha$ governs the
steepness of the cutoff in the spectrum. For low values of $\alpha$
we obtain spectra which exhibit a slowly varying cutoff rather than
the steep cutoff for $\alpha = 3.35$. This latter value mimics the
the CPKL fast-roll spectrum, and larger values than this would
approach the step function. Here we explore the softer cutoff
provided by small values of $\alpha$. As we did for the theta
function model, we keep the aspect of a gradient in the number of
e-folds of inflation. The resulting scenario may be relevant for
other models, with the basic features of an IR cutoff of the
perturbation spectrum with the scale of the cutoff varying spatially
as dictated by a gradient in the number of e-folds of inflation.

\begin{figure}
\begin{center}
  \begin{minipage}[t]{.07\textwidth}
    \vspace{0pt}
    \centering
    \vspace*{99pt}
    \hspace*{-10pt}
    \rotatebox{90}{$F_\ell^D$}
  \end{minipage}%
  \begin{minipage}[t]{0.83\textwidth}
    \vspace{0pt}
    \centering
    \includegraphics[width=0.99\textwidth,height=!]{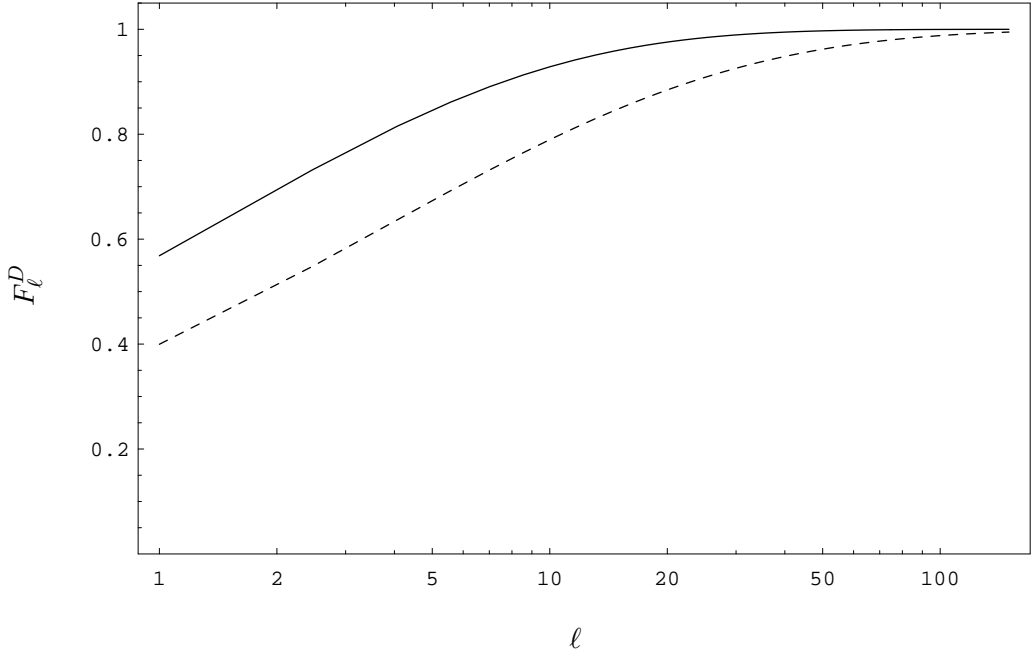}
  \end{minipage}
\end{center}
$\hspace*{240pt} \ell \vspace*{10pt}$ \caption{\small Suppression
functions resulting from averaged exponential spectra for the
northern (dashed) and southern (solid) hemisphere for
         $\bar \alpha = 0.5$, $\ell_0 = 2$ and $\Delta N_{low} = 2$. Note that the underlying calculation neglected the integrated Sachs-Wolfe effect. \vspace*{10pt}} \label{fig_NS_exp_sup}
\end{figure}

As we did in the case of the theta function spectrum analysis, we
do not include a spectral index so that Eq.
(\ref{exponentialcutoff}) simplifies to
\begin{equation}
 P_{\Phi}(k) = 1 -
 \exp \big[-(k/k_{c})^\alpha\big]. \label{eq_exponentialcutoff_nospecindex}
\end{equation}
We find that when one numerically calculates the Sachs-Wolfe
spectrum of such an exponential cutoff spectrum via Eq.
(\ref{eq_sachs_wolfe}), the resulting Sachs-Wolfe spectrum is
approximated extremely well by an exponential cutoff spectrum for
$\ell(\ell+1) \, C_\ell$ (where the exponential $\bar \alpha$ for
the Sachs-Wolfe spectrum is a bit smaller than $\alpha$ in
$P_{\Phi}(k)$). Ignoring again the integrated Sachs-Wolfe effect, we can
therefore use
\begin{equation}
 \ell(\ell+1) \, C_\ell = 1 -
 \exp \big[-(\ell/\ell_{c})^{\bar \alpha}\big] \label{eq_exponentialcutoff_cl}
\end{equation}
where now the position of the cutoff in $\ell$ space varies
spatially as
\begin{equation}
 \ell_c = \ell_0 \exp\left(+\Delta N_{low} \, \frac{\zeta}{\zeta_{low}}\right)
\end{equation}
so that
\begin{equation}
 C_\ell (\zeta) = \frac{1}{\ell(\ell+1)} \left(1 -
 \exp \left[-\left(\frac{\ell}{\ell_0 \exp\left(+\Delta N_{low} \, \frac{\zeta}{\zeta_{low}}\right)}\right)^{\bar \alpha}\right]\right). \label{eq_exponentialcutoff_cl_zeta}
\end{equation}
To find the averaged spectra for the two hemispheres, we integrate
over the position dependent spectrum analogously to Eq.
(\ref{eq_av_spec_hemis})
\begin{align} \label{eq_av_cl_hemis}
 C^N_\ell & = \int_0^{\zeta_{low}} d\zeta \, \exp\left(-\Delta N_{low} \, \frac{\zeta}{\zeta_{low}}\right) \, C_\ell (\zeta) \nonumber \\
 C^S_\ell & = \int_{\zeta_{high}}^0 d\zeta \, \exp\left(-\Delta N_{low} \, \frac{\zeta}{\zeta_{low}}\right) \, C_\ell (\zeta).
\end{align}

For the parameters $\bar \alpha = 0.5$, $\ell_0 = 2$ and $\Delta
N_{low} = 2$, the suppression functions $F_\ell^D$ for $D = N,S$
are displayed in Fig. \ref{fig_NS_exp_sup}. From this we see that
it is possible to have a suppression factor that extends out to
larger values of $\ell$, and which creates an asymmetry on the
sky. Treated simply as a suppression of the standard WMAP
spectrum, this modification will not lead to an improved fit
because the suppression of the power spectrum occurs in both
hemispheres. It is possible that a better fit could be obtained if
one adjusts the magnitude of the power spectrum, but this would
involve adjustments of all the astrophysical parameters in order
to not upset the agreement obtained at smaller angular scales.


\begin{thebibliography}{99} \itemsep -1pt

\bibitem{Contaldi:2003zv}
  C.~R.~Contaldi, M.~Peloso, L.~Kofman and A.~Linde,
  ``Suppressing the lower Multipoles in the CMB Anisotropies,''
  JCAP {\bf 0307}, 002 (2003)
  [arXiv:astro-ph/0303636].

\bibitem{inflation}
  A.~H.~Guth,
  ``The Inflationary Universe: A Possible Solution To The Horizon And Flatness
  Problems,''
  Phys.\ Rev.\ D {\bf 23}, 347 (1981).\\
  A.~D.~Linde,
  ``A New Inflationary Universe Scenario: A Possible Solution Of The Horizon,
  Flatness, Homogeneity, Isotropy And Primordial Monopole Problems,''
  Phys.\ Lett.\ B {\bf 108}, 389 (1982).\\
  A.~Albrecht and P.~J.~Steinhardt,
  ``Cosmology For Grand Unified Theories With Radiatively Induced Symmetry
  Breaking,''
  Phys.\ Rev.\ Lett.\  {\bf 48}, 1220 (1982).\\
For details, see A. D. Linde, {\it Particle Physics and
Inflationary Cosmology}, (Harwood, Chur, Switzerland, 1990)
[arXiv:hep-th/0503203].

\bibitem{Spergel:2006hy}
  D.~N.~Spergel {\it et al.},
  ``Wilkinson Microwave Anisotropy Probe (WMAP) three year results:
  Implications for cosmology,''
  arXiv:astro-ph/0603449.
\bibitem{Eriksen}
 H.~K.~Eriksen, F.~K.~Hansen, A.~J.~Banday, K.~M.~Gorski and P.~B.~Lilje,
  ``Asymmetries in the CMB anisotropy field,''
  Astrophys.\ J.\  {\bf 605}, 14 (2004)
  [Erratum-ibid.\  {\bf 609}, 1198 (2004)]
  [arXiv:astro-ph/0307507].
\bibitem{Hansen:2004vq}
  F.~K.~Hansen, A.~J.~Banday and K.~M.~Gorski,
  ``Testing the cosmological principle of isotropy: local power spectrum
  estimates of the WMAP data,''   Mon.\ Not.\ Roy.\ Astron.\ Soc.\  {\bf 354}, 641 (2004)
  [arXiv:astro-ph/0404206].


\bibitem{north_south_non_gaussianity}

  C.~G.~Park,
  ``Non-Gaussian Signatures in the Temperature Fluctuation Observed by the
  WMAP,''
  Mon.\ Not.\ Roy.\ Astron.\ Soc.\  {\bf 349}, 313 (2004)
  [arXiv:astro-ph/0307469].

    H.~K.~Eriksen, D.~I.~Novikov, P.~B.~Lilje, A.~J.~Banday and K.~M.~Gorski,
  ``Testing for non-Gaussianity in the WMAP data: Minkowski functionals and the
  length of the skeleton,''
  Astrophys.\ J.\  {\bf 612}, 64 (2004)
  [arXiv:astro-ph/0401276].
\bibitem{general_anomaly}
  C.~Gordon,
  ``Broken Isotropy from a Linear Modulation of the Primordial Perturbations,''
  arXiv:astro-ph/0607423.\\
  E.~Komatsu {\it et al.},
  ``First Year Wilkinson Microwave Anisotropy Probe (WMAP) Observations: Tests
  of Gaussianity,''
  Astrophys.\ J.\ Suppl.\  {\bf 148}, 119 (2003)
  [arXiv:astro-ph/0302223].\\
  P.~Coles, P.~Dineen, J.~Earl and D.~Wright,
  ``Phase Correlations in Cosmic Microwave Background Temperature Maps,''
  Mon.\ Not.\ Roy.\ Astron.\ Soc.\  {\bf 350}, 983 (2004)
  [arXiv:astro-ph/0310252].\\
  P.~Mukherjee and Y.~Wang,
  ``Wavelets and WMAP non-Gaussianity,''
  Astrophys.\ J.\  {\bf 613}, 51 (2004)
  [arXiv:astro-ph/0402602].\\
  K.~Land and J.~Magueijo,
 ``The axis of evil,''
  Phys.\ Rev.\ Lett.\  {\bf 95}, 071301 (2005)
  [arXiv:astro-ph/0502237].\\
  A.~de Oliveira-Costa, M.~Tegmark, M.~Zaldarriaga and A.~Hamilton,
 ``The significance of the largest scale CMB fluctuations in WMAP,''
  Phys.\ Rev.\ D {\bf 69}, 063516 (2004)
  [arXiv:astro-ph/0307282].\\
  P.~Coles, P.~Dineen, J.~Earl and D.~Wright,
  ``Phase Correlations in Cosmic Microwave Background Temperature Maps,''
  Mon.\ Not.\ Roy.\ Astron.\ Soc.\  {\bf 350}, 983 (2004)
  [arXiv:astro-ph/0310252].\\
  D.~J.~Schwarz, G.~D.~Starkman, D.~Huterer and C.~J.~Copi,
  ``Is the low-l microwave background cosmic?,''
  Phys.\ Rev.\ Lett.\  {\bf 93}, 221301 (2004)
  [arXiv:astro-ph/0403353].\\
   C.~J.~Copi, D.~Huterer, D.~J.~Schwarz and G.~D.~Starkman,
  ``On the large-angle anomalies of the microwave sky,''
  Mon.\ Not.\ Roy.\ Astron.\ Soc.\  {\bf 367}, 79 (2006)
  [arXiv:astro-ph/0508047].\\
   C.~J.~Copi, D.~Huterer and G.~D.~Starkman,
  ``Multipole Vectors--a new representation of the CMB sky and evidence for
  statistical anisotropy or non-Gaussianity at $2 \leq l \leq 8$,''
  Phys.\ Rev.\ D {\bf 70}, 043515 (2004)
  [arXiv:astro-ph/0310511].\\
    Y.~Wiaux, P.~Vielva, E.~Martinez-Gonzalez and P.~Vandergheynst,
  ``Global universe anisotropy probed by the alignment of structures in the
  cosmic microwave background,''
  Phys.\ Rev.\ Lett.\  {\bf 96}, 151303 (2006)
  [arXiv:astro-ph/0603367].
\bibitem{coldspot}
  D.~L.~Larson and B.~D.~Wandelt,
  ``The Hot and Cold Spots in the WMAP Data are Not Hot and Cold Enough,''
  Astrophys.\ J.\  {\bf 613}, L85 (2004)
  [arXiv:astro-ph/0404037].\\
    M.~Cruz, E.~Martinez-Gonzalez, P.~Vielva and L.~Cayon,
  ``Detection of a non-Gaussian Spot in WMAP,''
  Mon.\ Not.\ Roy.\ Astron.\ Soc.\  {\bf 356}, 29 (2005)
  [arXiv:astro-ph/0405341].\\
  P.~Vielva, E.~Martinez-Gonzalez, R.~B.~Barreiro, J.~L.~Sanz and L.~Cayon,
  ``Detection of non-Gaussianity in the WMAP 1-year data using spherical
  wavelets,''
  Astrophys.\ J.\  {\bf 609}, 22 (2004)
  [arXiv:astro-ph/0310273].


\bibitem{Eriksen:2007pc}
  H.~K.~Eriksen, A.~J.~Banday, K.~M.~Gorski, F.~K.~Hansen and P.~B.~Lilje,
  ``Hemispherical power asymmetry in the three-year Wilkinson Microwave
  Anisotropy Probe sky maps,''
  arXiv:astro-ph/0701089.

  \bibitem{collection}
 C.~Copi, D.~Huterer, D.~Schwarz and G.~Starkman,
  ``The Uncorrelated Universe: Statistical Anisotropy and the Vanishing Angular
  Correlation Function in WMAP Years 1-3,''
  arXiv:astro-ph/0605135.\\
 T.~R.~Jaffe, A.~J.~Banday, H.~K.~Eriksen, K.~M.~Gorski and F.~K.~Hansen,
  ``Evidence of vorticity and shear at large angular scales in the WMAP  data:
  A violation of cosmological isotropy?,''
  Astrophys.\ J.\  {\bf 629}, L1 (2005)
  [arXiv:astro-ph/0503213].\\
 K.~Land and J.~Magueijo,
  ``The Axis of Evil revisited,''
  arXiv:astro-ph/0611518.


\bibitem{Hansen:2006rj}
  F.~K.~Hansen, A.~J.~Banday, H.~K.~Eriksen, K.~M.~Gorski and P.~B.~Lilje,
  ``Foreground Subtraction of Cosmic Microwave Background Maps using WI-FIT
  (Wavelet based hIgh resolution Fitting of Internal Templates),''
  arXiv:astro-ph/0603308.

\bibitem{Maino:2006pq}
  D.~Maino, S.~Donzelli, A.~J.~Banday, F.~Stivoli and C.~Baccigalupi,
  ``CMB signal in WMAP 3yr data with FastICA,''
  arXiv:astro-ph/0609228.

\bibitem{Raeth:2007ti}
  C.~Raeth, P.~Schuecker and A.~J.~Banday,
  ``A Scaling Index Analysis of the WMAP three year data: Signatures of
  non-Gaussianities and Asymmetries in the CMB,''
  arXiv:astro-ph/0702163.

\bibitem{wmap5_evidence}
  J.~Hoftuft, H.~K.~Eriksen, A.~J.~Banday, K.~M.~Gorski, F.~K.~Hansen and P.~B.~Lilje,
  ``Increasing evidence for hemispherical power asymmetry in the five-year WMAP
  data,''
  arXiv:0903.1229 [astro-ph.CO].\\
  F.~K.~Hansen, A.~J.~Banday, K.~M.~Gorski, H.~K.~Eriksen and P.~B.~Lilje,
  ``Power Asymmetry in Cosmic Microwave Background Fluctuations from Full Sky
  to Sub-degree Scales: Is the Universe Isotropic?,''
  arXiv:0812.3795 [astro-ph].


\bibitem{Hajian}
  A.~Hajian,
  ``Analysis of the apparent lack of power in the cosmic microwave background
  anisotropy at large angular scales,''
  arXiv:astro-ph/0702723.\\
  A.~Hajian, T.~Souradeep and N.~J.~Cornish,
  ``Statistical Isotropy of the WMAP Data: A Bipolar Power Spectrum Analysis,''
  Astrophys.\ J.\  {\bf 618}, L63 (2004)
  [arXiv:astro-ph/0406354].


\bibitem{evan}
 E.~P.~Donoghue and J.~F.~Donoghue,
  ``Isotropy of the early universe from CMB anisotropies,''
  Phys.\ Rev.\ D {\bf 71}, 043002 (2005)
  [arXiv:astro-ph/0411237].


\bibitem{Gumrukcuoglu:2006xj}
  A.~E.~Gumrukcuoglu, C.~R.~Contaldi and M.~Peloso,
  arXiv:astro-ph/0608405. \\
  A.~E.~Gumrukcuoglu, C.~R.~Contaldi and M.~Peloso,
  JCAP {\bf 0711}, 005 (2007)
  [arXiv:0707.4179 [astro-ph]].


\bibitem{models}
  C.~Vale,
  ``Local Pancake Defeats Axis of Evil,''
  arXiv:astro-ph/0509039.\\
   R.~V.~Buniy,
  ``Eccentric inflation and WMAP,''
  Int.\ J.\ Mod.\ Phys.\ A {\bf 20}, 1095 (2005)
  [arXiv:hep-ph/0408026].\\
  D.~Langlois and T.~Piran,
  ``Dipole anisotropy from an entropy gradient,''
  Phys.\ Rev.\ D {\bf 53}, 2908 (1996)
  [arXiv:astro-ph/9507094].\\
  S.~H.~S.~Alexander,
  ``Is cosmic parity violation responsible for the anomalies in the WMAP
  data?,''
  arXiv:hep-th/0601034.\\
  K.~Land and J.~Magueijo,
  ``Template fitting and the large-angle CMB anomalies,''
  [arXiv:astro-ph/0509752].\\
   B.~Feng and X.~Zhang,
  ``Double inflation and the low CMB quadrupole,''
  Phys.\ Lett.\ B {\bf 570}, 145 (2003)
  [arXiv:astro-ph/0305020].\\
  J.~Levin,
  ``Topology and the cosmic microwave background,''
  Phys.\ Rept.\  {\bf 365}, 251 (2002)
  [arXiv:gr-qc/0108043].\\
  A.~de Oliveira-Costa, G.~F.~Smoot and A.~A.~Starobinsky,
  ``Can the lack of symmetry in the COBE/DMR maps constrain the topology of the
  universe?,''
  Astrophys.\ J.\  {\bf 468}, 457 (1996)
  [arXiv:astro-ph/9510109].\\
  L.~Ackerman, S.~M.~Carroll and M.~B.~Wise,
  ``Imprints of a Primordial Preferred Direction on the Microwave Background,''
  arXiv:astro-ph/0701357.\\
  T.~Koivisto and D.~F.~Mota,
  ``Accelerating Cosmologies with an Anisotropic Equation of State,''
  Astrophys.\ J.\  {\bf 679}, 1 (2008)
  [arXiv:0707.0279 [astro-ph]].\\
  C.~G.~Boehmer and D.~F.~Mota,
  ``CMB Anisotropies and Inflation from Non-Standard Spinors,''
  Phys.\ Lett.\  B {\bf 663}, 168 (2008)
  [arXiv:0710.2003 [astro-ph]].\\
  A.~L.~Erickcek, M.~Kamionkowski and S.~M.~Carroll,
  Phys.\ Rev.\  D {\bf 78}, 123520 (2008)
  [arXiv:0806.0377 [astro-ph]].\\
  A.~L.~Erickcek, S.~M.~Carroll and M.~Kamionkowski,
  ``Superhorizon Perturbations and the Cosmic Microwave Background,''
  Phys.\ Rev.\  D {\bf 78}, 083012 (2008)
  [arXiv:0808.1570 [astro-ph]].\\
  C.~Pitrou, T.~S.~Pereira and J.~P.~Uzan,
  ``Predictions from an anisotropic inflationary era,''
  JCAP {\bf 0804}, 004 (2008)
  [arXiv:0801.3596 [astro-ph]].\\
  A.~R.~Pullen and M.~Kamionkowski,
  ``Cosmic Microwave Background Statistics for a Direction-Dependent Primordial
  Power Spectrum,''
  Phys.\ Rev.\  D {\bf 76}, 103529 (2007)
  [arXiv:0709.1144 [astro-ph]].\\
  K.~Dimopoulos, M.~Karciauskas, D.~H.~Lyth and Y.~Rodriguez,
  ``Statistical anisotropy of the curvature perturbation from vector field
  perturbations,''
  arXiv:0809.1055 [astro-ph].\\
  X.~Gao,
  ``Can Relic Superhorizon Inhomogeneities be Responsible for Large-Scale CMB
  Anomalies?,''
  arXiv:0903.1412 [astro-ph.CO].\\
  Y.~Shtanov and H.~Pyatkovska,
  ``Statistical anisotropy in the inflationary universe,''
  arXiv:0904.1887 [gr-qc].


\bibitem{Explantion_north_south}
  J.~W.~Moffat,
  ``Cosmic Microwave Background, Accelerating Universe and Inhomogeneous
  Cosmology,''
  JCAP {\bf 0510}, 012 (2005)
  [arXiv:astro-ph/0502110].\\
  C.~Gordon, W.~Hu, D.~Huterer and T.~Crawford,
  ``Spontaneous Isotropy Breaking: A Mechanism for CMB Multipole Alignments,''
  [arXiv:astro-ph/0509301].\\
  K.~T.~Inoue and J.~Silk,
  ``Local Voids as the Origin of Large-angle Cosmic Microwave Background
  Anomalies: The Effect of a Cosmological Constant,''
  arXiv:astro-ph/0612347.

\bibitem{other_suppression}
 B.~A.~Powell and W.~H.~Kinney,
  ``The pre-inflationary vacuum in the cosmic microwave background,''
  arXiv:astro-ph/0612006.\\
 L.~Covi, J.~Hamann, A.~Melchiorri, A.~Slosar and I.~Sorbera,
  ``Inflation and WMAP three year data: Features have a future!,''
  Phys.\ Rev.\ D {\bf 74}, 083509 (2006)
  [arXiv:astro-ph/0606452].\\
   D.~Langlois and F.~Vernizzi,
  ``From heaviness to lightness during inflation,''
  JCAP {\bf 0501}, 002 (2005)
  [arXiv:astro-ph/0409684].

   \bibitem{extrawork}
  J.~M.~Cline, P.~Crotty and J.~Lesgourgues,
  ``Does the small CMB quadrupole moment suggest new physics?,''
  JCAP {\bf 0309}, 010 (2003)
  [arXiv:astro-ph/0304558].\\
  R.~Sinha and T.~Souradeep,
  ``Post-WMAP assessment of infrared cutoff in the primordial spectrum from
  inflation,''
  Phys.\ Rev.\ D {\bf 74}, 043518 (2006)
  [arXiv:astro-ph/0511808].

\bibitem{Mukhanov:2005sc}
  V.~Mukhanov,
  {\it Physical foundations of cosmology} (Cambridge University Press,
  2005).

\bibitem{Mukhanov:1990me}
  V.~F.~Mukhanov, H.~A.~Feldman and R.~H.~Brandenberger,
  ``Theory of cosmological perturbations. Part 1. Classical perturbations. Part
  2. Quantum theory of perturbations. Part 3. Extensions,''
  Phys.\ Rept.\  {\bf 215}, 203 (1992).

\bibitem{CMBFAST}
  U.~Seljak and M.~Zaldarriaga,
  Astrophys.\ J.\  {\bf 469}, 437 (1996)
  [arXiv:astro-ph/9603033],
  \ http://www.cmbfast.org/

\bibitem{CMBEASY}
  M.~Doran,
  JCAP {\bf 0510}, 011 (2005)
  [arXiv:astro-ph/0302138],
  \ http://www.cmbeasy.org/

\bibitem{CAMB}
  A.~Lewis, A.~Challinor and A.~Lasenby,
  Astrophys.\ J.\  {\bf 538}, 473 (2000)
  [arXiv:astro-ph/9911177],
  \ http://camb.info/

\bibitem{Finelli:2005zc}
  F.~Finelli and A.~Gruppuso,
  arXiv:hep-th/0501089.

\bibitem{Hinshaw:2006ia}
  G.~Hinshaw {\it et al.}  [WMAP Collaboration],
  ``Three-year Wilkinson Microwave Anisotropy Probe (WMAP) observations:
  Temperature analysis,''
  arXiv:astro-ph/0603451.


\bibitem{Sachs:1967er}
  R.~K.~Sachs and A.~M.~Wolfe,
  ``Perturbations of a cosmological model and angular variations of the
  microwave background,''
  Astrophys.\ J.\  {\bf 147}, 73 (1967).

\bibitem{LiddleandLyth}
  A.R~Liddle, D.H~Lyth,
  {\it Cosmological Inflation and Large-Scale structure} (Cambridge University Press,
  2000).



\end{thebibliography}
\end{document}